\documentclass[%
reprint,
superscriptaddress,
 amsmath,amssymb,
 aps,
]{revtex4-2}

\usepackage{graphicx}
\usepackage{subcaption} 
\DeclareUnicodeCharacter{2061}{}
\usepackage{dcolumn}
\usepackage{bm}
\usepackage[usenames,dvipsnames]{color} 
\usepackage{soul}

\begin{document}
\preprint{APS/123-QED}
\title{Generation of multiple ultrashort solitons in a third-order nonlinear composite medium with self-focusing and self-defocusing nonlinearities}

\author{Andr\'e C. A. Siqueira}
\thanks{Corresponding author: andrechaves.physics@gmail.com}%
\author{Edilson L. Falc\~ao-Filho}%

\affiliation{Departamento de F\'isica, Universidade Federal de Pernambuco, 50670-901 Recife, PE, Brazil
}%

\author{Boris A. Malomed}
\affiliation{Department of Physical Electronics, School of Electrical Engineering, Faculty of Engineering,
Tel Aviv University, Tel Aviv 69978, Israel
}
\affiliation{Instituto de Alta Investigaci\'on, Universidad de Tarapac\'a, Casilla 7D, Arica, Chile}
\author{Cid B. de Ara\'ujo}
\affiliation{Departamento de F\'isica, Universidade Federal de Pernambuco, 50670-901 Recife, PE, Brazil
}


\date{\today}

\begin{abstract}

Theoretical consideration of the propagation of femtosecond-Gaussian pulses in a 1D composite medium, consisting of alternating self-focusing (SF) and self-defocusing (SDF) waveguide segments with normal group-velocity dispersion predicts the generation of trains of bright solitons when an optical pulse first propagates in the SF segment, followed by the SDF one. The multiple temporal compression (MTC) process, based on this setting, offers a method for controllable generation of multiple ultrashort temporal solitons. Numerical solutions of the generalized nonlinear Schr\"{o}dinger equation modeling this system demonstrate that the intrapulse Raman scattering plays a major role in the temporal and spectral dynamics. Collisions between ultrashort solitons with different central wavelengths are addressed too. The paper provides, for the first time, a procedure for producing controllable trains of ultrashort temporal solitons by incident optical pulses propagating in a composite medium.

\end{abstract}

\maketitle

\section{\label{sec:level1}Introduction}
The dynamics of pulse propagation in media with anomalous dispersion and self-focusing nonlinearity is crucially important for soliton formation. Indeed, considering higher-order dispersion and the nonlinear (NL) terms affecting the pulse propagation, such as the self-phase modulation, self-steepening, and intrapulse Raman scattering, the generation of multiple solitons may occur during the pulse propagation, initiated by temporal compression followed by soliton fission
\cite{sysoliatin2007soliton, tai1988fission, driben2013newton, dudley2006supercontinuum,dudley2002numerical}. Most commonly, the input pulse evolves into a single temporal soliton with the remaining energy being spilled out as dispersive waves. However, under specific conditions, the energy released by the first soliton fission may also evolve into multiple temporal solitons \cite{dudley2006supercontinuum,dudley2002numerical, braud2016solitonization, demircan2008effects, bose2015experimental}. Then, as the first fission takes place around the central peak of the input pulse, the first emerging soliton has a higher peak power than the secondary solitons. Hence, the first solitary pulse features a stronger soliton self-frequency shift (SSFS), which is accompanied by temporal deceleration through the intrapulse Raman scattering and anomalous dispersion \cite{dudley2006supercontinuum,dudley2002numerical, demircan2008effects, bose2015experimental, braud2016solitonization, gordon1986theory, mitschke1986discovery, Agrawal2013nonlinear}.

Also, efforts have been made to explore media providing a negative NL refractive index (\(n_{2}\)) in the normal-dispersion regime. For instance, birefringent NL crystals, such as LiNbO$_{3}$, BBO, and KTP, have been investigated to control the effective NL refractive index ($n_{2,eff}$) acting on the pulse, where the strong negative second-order cascading NL regime can overcompensate the positive Kerr nonlinearity; then, a medium with a defocusing effective nonlinearity, \(n_{2,eff}< 0\), in the normal dispersion regime \cite{desalvo1992self, ashihara2002soliton, bache2008limits} is obtained. Such media have also been used to compress pulses, and to control the supercontinuum generation \cite{guo2014few, vsuminas2017second, conforti2013extreme}. 

On the other hand, studies of composites containing metallic nanoparticles (NPs) have drawn much interest of the NL optics community. These metal-dielectric nanocomposites may present large and controllable $n_{2,eff}$ due to contributions of the host material and NPs. The amplitude and sign of the nonlinearity can be controlled by proper selection of size, shape, and volume fraction of the NPs \cite{zhang2017nonlinear, reyna2017high, kassab2018metal, reyna2022beyond}. For example, it was shown that optical fibers based on fused silica, doped with silver NPs, exhibit $n_{2,eff} < 0$, and, in the case of the normal dispersion, they can provide the generation of bright solitons controlled by the volume fraction of the metal NPs and the central wavelength of the pulse \cite{bose2016study, arteaga2018soliton, bose2018dispersive, bose2016implications, driben2010solitary, zhao2022effects}.

In the present work, we report numerical results, based on the generalized NL Schr\"{o}dinger equation (GNLSE), which exhibits generation of trains of ultrashort solitons under the action of the normal dispersion, without soliton fission near the central peak of the pulse. Our approach addresses the propagation of 800 nm femtosecond pulses in a stacked 1D system in such a way that the pulse, initially, propagates in a self-focusing (SF) waveguide followed by a self-defocusing (SDF) one. The dispersion of both waveguide segments is assumed to be normal in the entire composite system, providing generation of solitons via what is called the \emph{multiple temporal compression} (MTC) process. For the calculation, the first segment in the waveguide is assumed to be made of pure fused silica (with $n_{2} > 0$), while the second one is made of silica doped with silver NPs, having $n_{2} < 0$. In this scenario, all generated solitons feature a temporal acceleration due to the intrapulse Raman scattering.

\section{\label{sec:level2}The theoretical approach }

To perform simulations of the pulse propagation through the two segments of the composite medium with $n_2$ of opposite signs, pure and doped fused silica based waveguides were considered.

The unidimensional propagation of an optical pulse, represented by a slowly varying envelope amplitude \(A(z,T)\) of the electric field, can be modeled by the GNLSE \cite{Agrawal2013nonlinear}, 

\begin{equation}
    \begin{split}
    \frac{\partial A}{\partial z} - \left(\sum\limits_{n\geqslant 2}\beta_{n}\frac{i^{n+1}}{n!}\frac{\partial^{n} A}{\partial T^{n}}\right) = i\gamma_{eff}\left(1 + \frac{i}{\omega_{0}} \frac{\partial}{\partial T}\right)( \left(1- f_{r}\right) \\
   A|A|^{2} + \frac{\gamma_{0}}{\gamma_{eff}}f_{r}A\int_{0}^{\infty}h_{r}(\tau)|A(z,T- \tau)|^{2}d\tau).
  \label{eq:1}
 \end{split}
\end{equation}
 
\begin{figure*}
    \centering
    \begin{subfigure}[b]{0.3\textwidth}
     \centering
        \includegraphics[width=\textwidth]{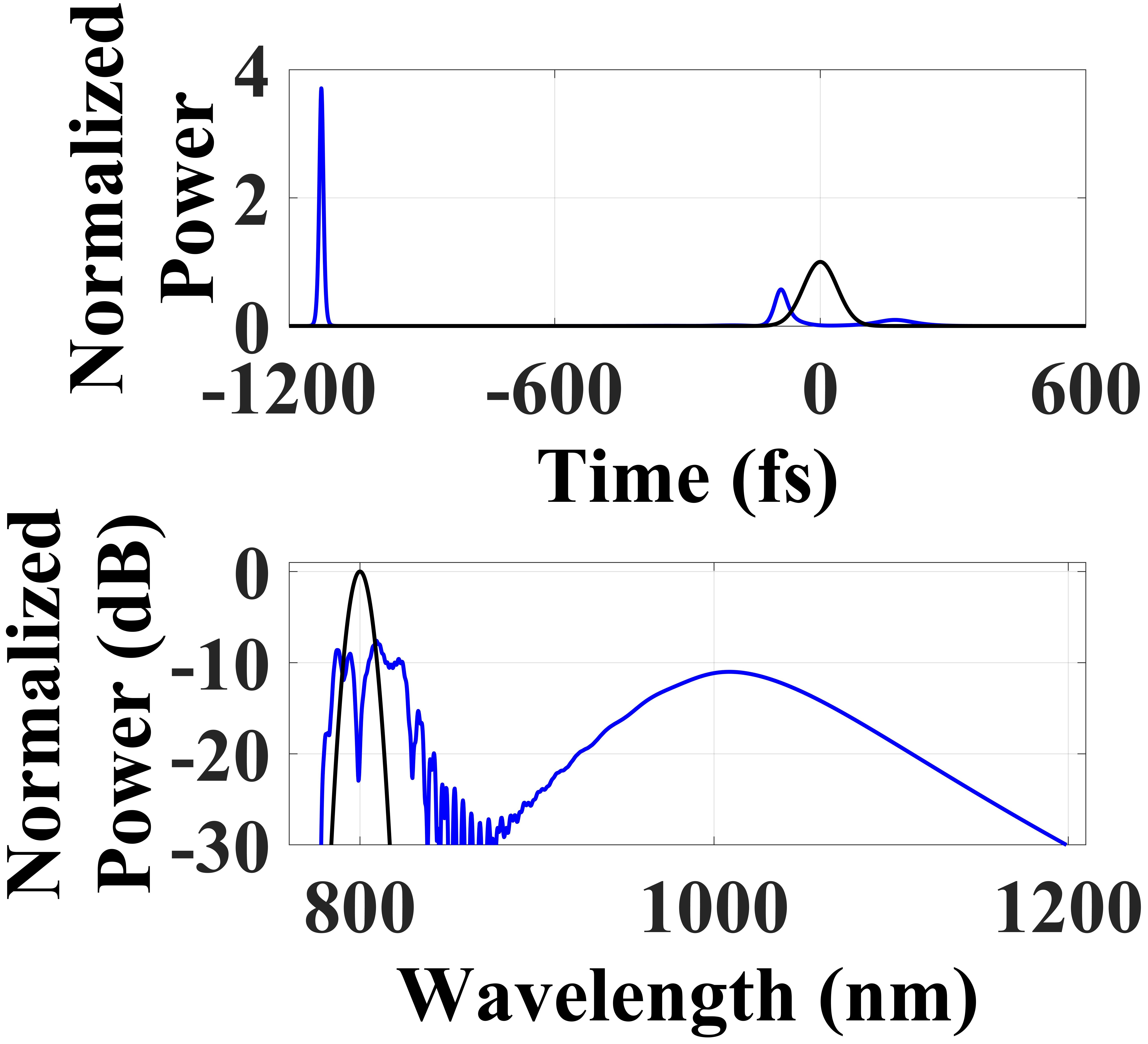}
         \caption{}
        \label{fig:1a}
    \end{subfigure}
    ~ 
    \begin{subfigure}[b]{0.315\textwidth}
     \centering
        \includegraphics[width=\textwidth]{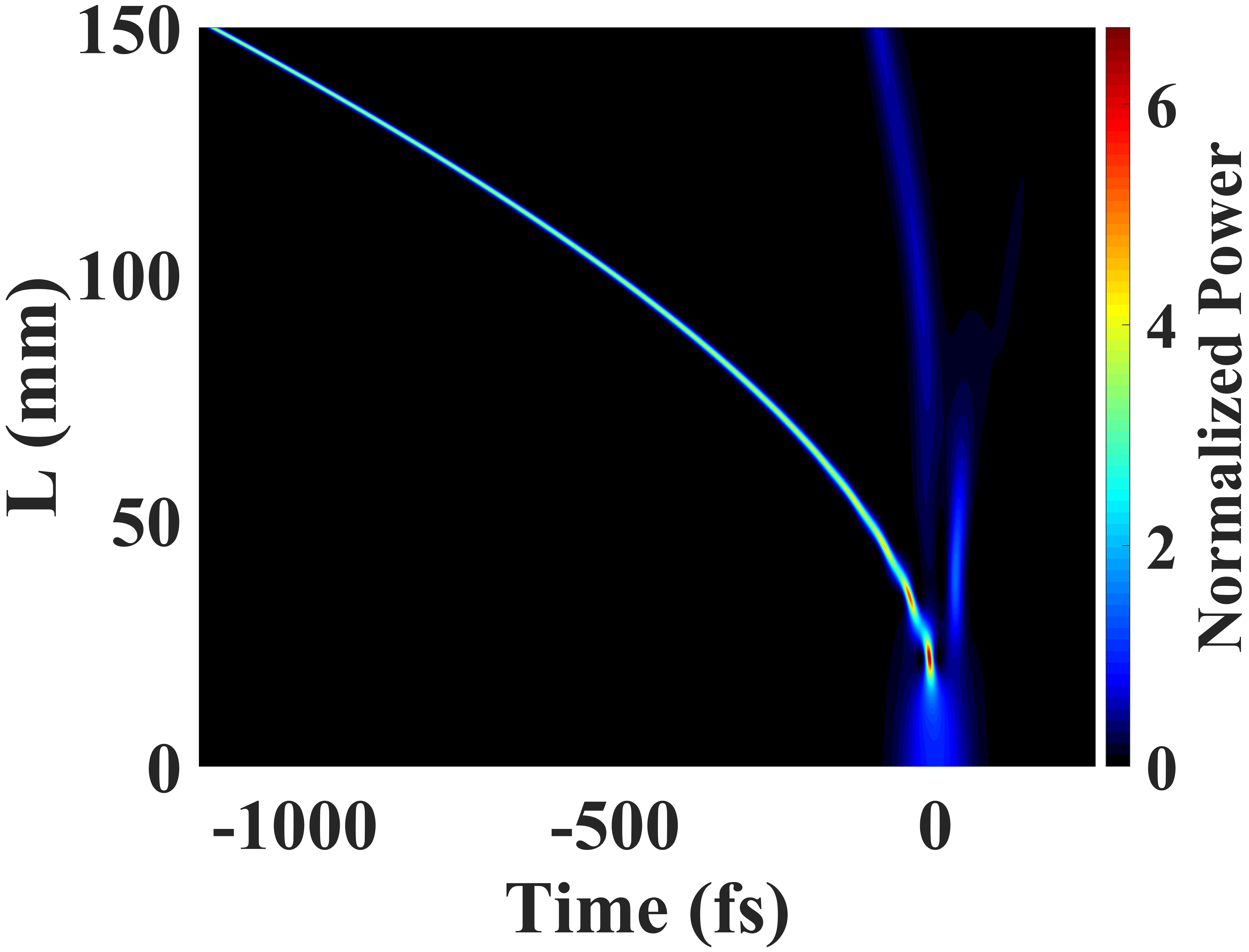}
         \caption{}
        \label{fig:1b}
    \end{subfigure}
    ~ 
    \begin{subfigure}[b]{0.32\textwidth}
     \centering
        \includegraphics[width=\textwidth]{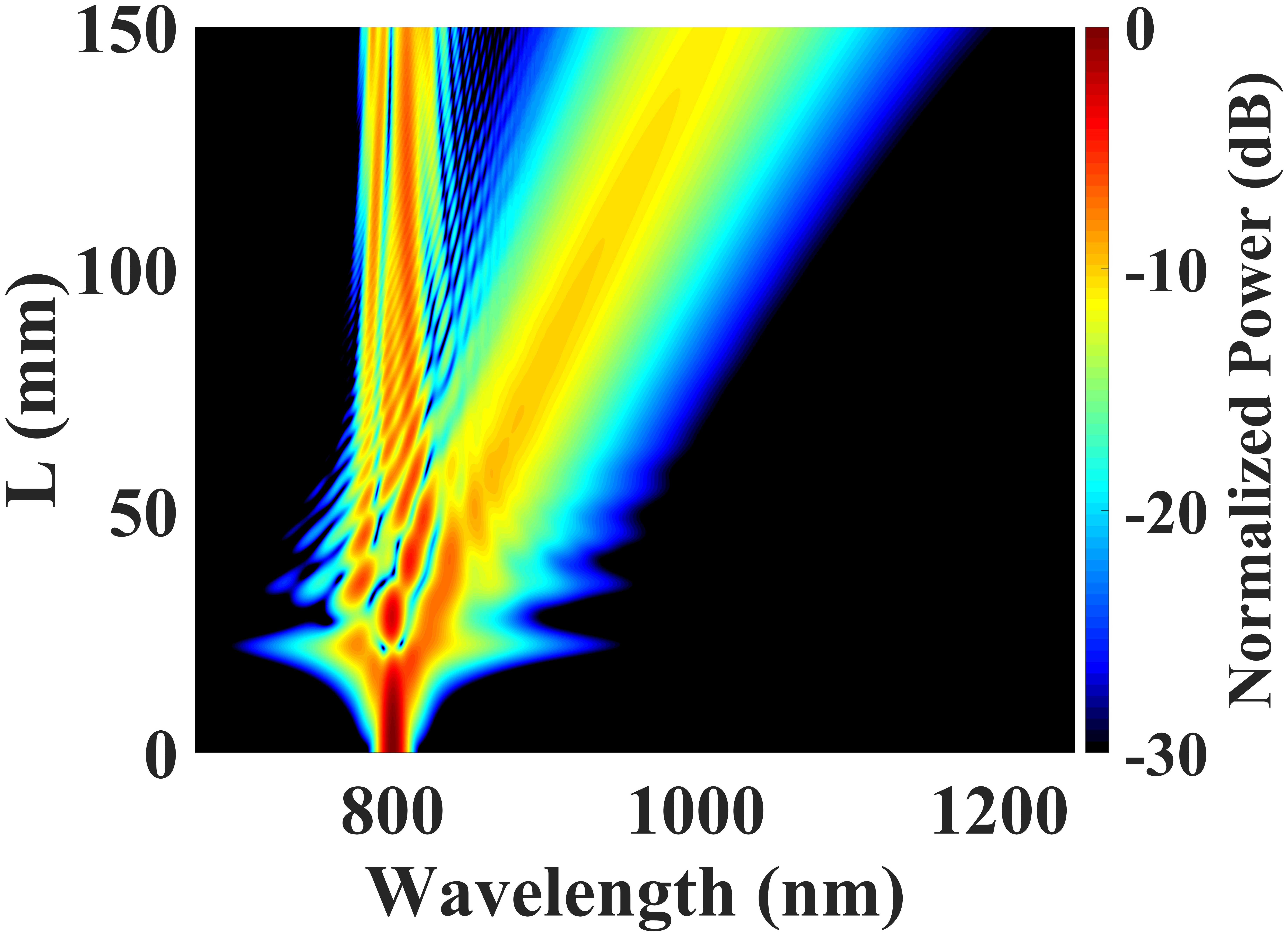}
         \caption{}
        \label{fig:1c}
    \end{subfigure}
    \begin{subfigure}[b]{0.32\textwidth}
     \centering
        \includegraphics[width=\textwidth]{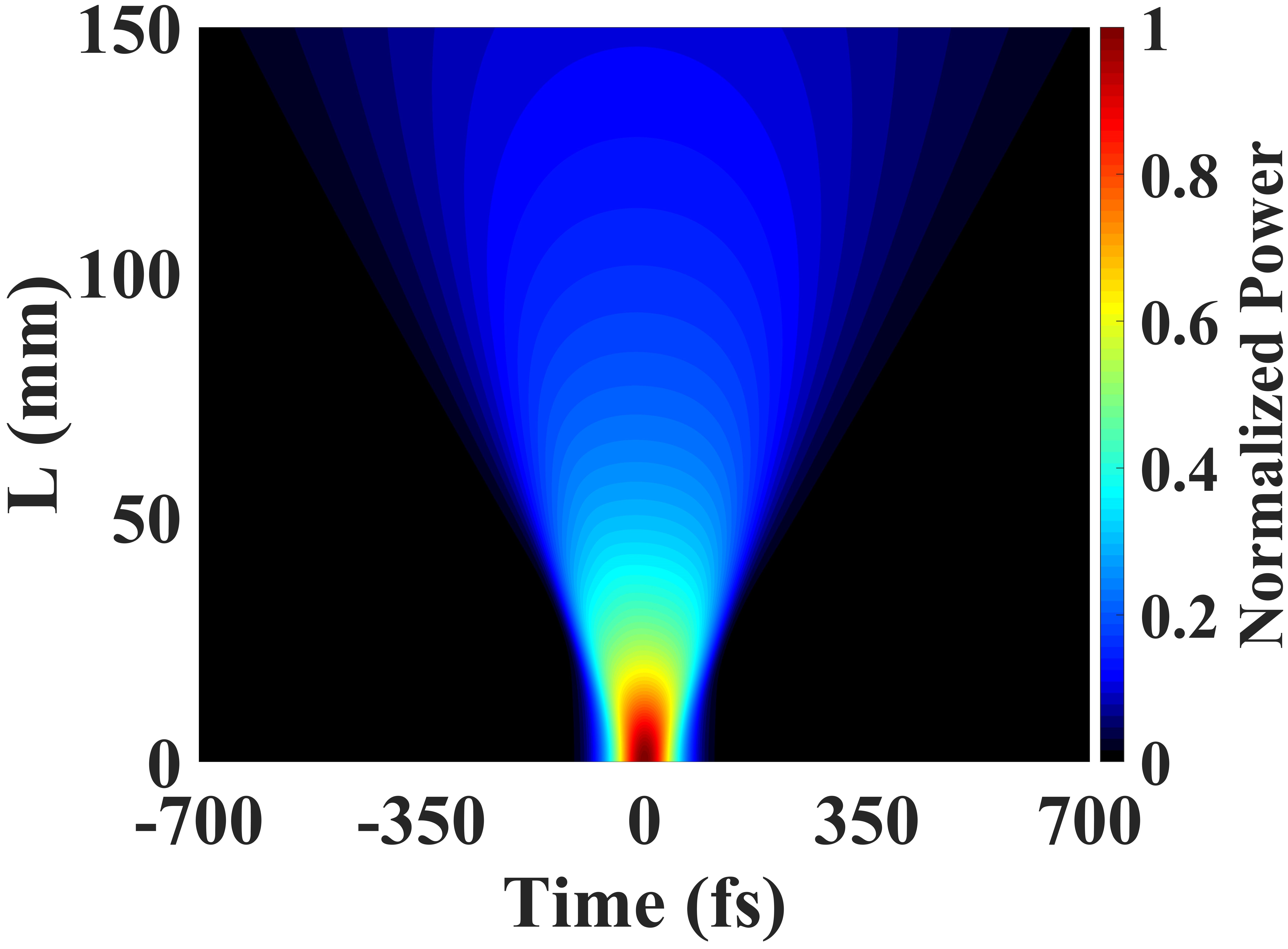}
         \caption{}
        \label{fig:1d}
    \end{subfigure}
  \begin{subfigure}[b]{0.32\textwidth}
        \includegraphics[width=\textwidth]{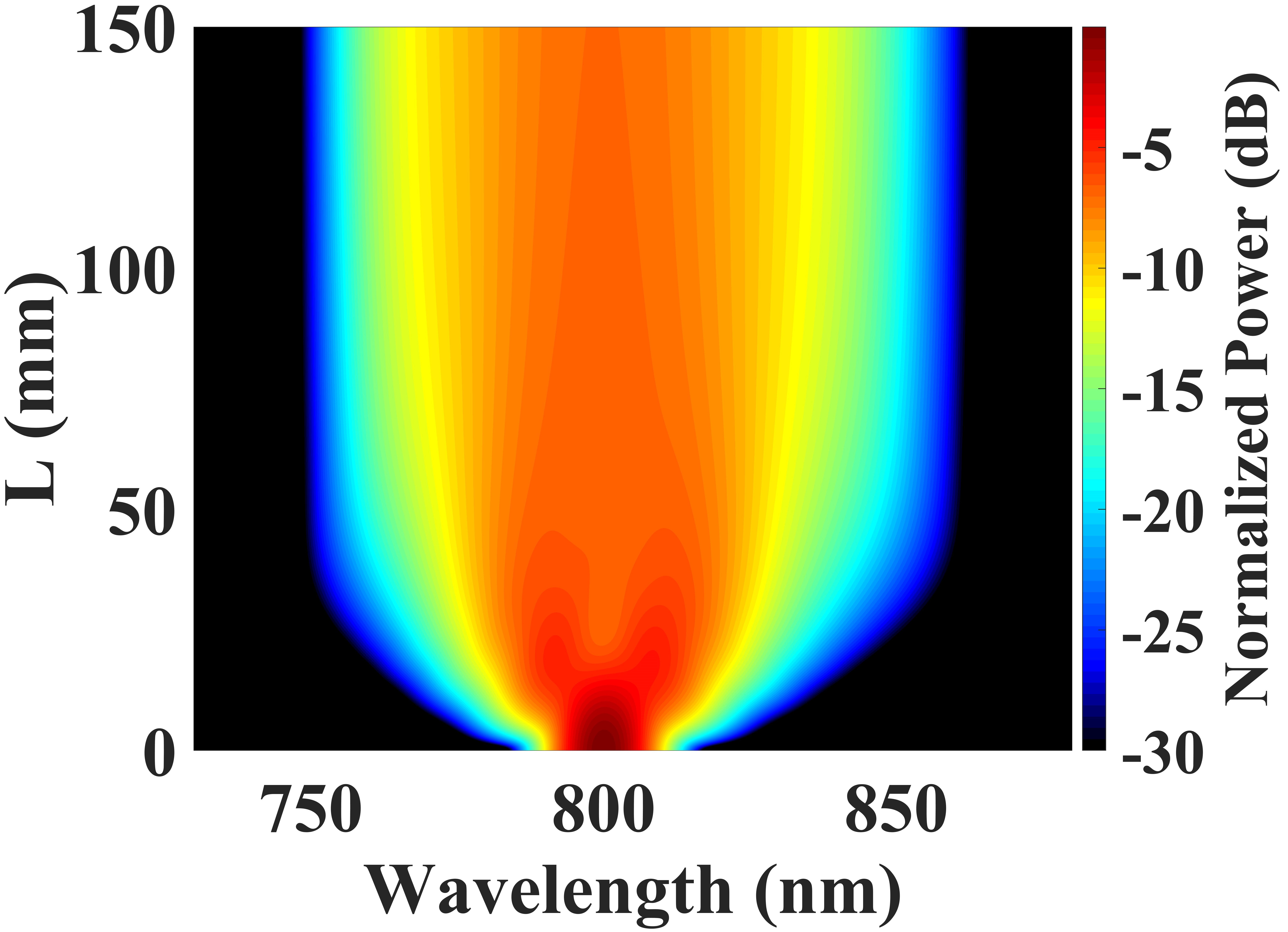}
         \caption{}
        \label{fig:1e}
    \end{subfigure}

    \caption{The soliton generation in the uniform medium under the action of the normal dispersion and SDF nonlinearity: (a) Temporal and spectral profiles of the input pulse (the black curve) and one obtained after passing 150 mm (the blue curve). (b,c) The respective temporal and spectral evolution, respectively. (d,e) The temporal and spectral evolution of the pulse in the medium with the normal dispersion and SF nonlinearity.}
    \label{fig:1}
\end{figure*}

Numerical solutions of Eq.(1) were performed by means of the fourth-order Runge-Kutta method in the interaction picture, which is more accurate in comparison to other methods, such as the conventional split-step Fourier-transform scheme \cite{hult2007fourth}. In Eq.(1), \(\beta_{n}\) (\(n\geqslant{2}\)) are the higher-order dispersion coefficients which come from the Taylor expansion of the propagation constant \(\beta(\omega)\) around the central frequency (\(\omega_{0}\)). The respective values of the high-order dispersion coefficients at 800 nm, numerically calculated from the Sellmeier expression for fused silica, are \cite{malitson1965interspecimen}: \(\beta_{2} = +3.63\cdot 10^{-2} ps^{2}m^{-1}\) , \(\beta_{3} =+ 2.75\cdot 10^{-5} ps^{3}m^{-1}\), \(\beta_{4} = -1.10\cdot 10^{-8} ps^{4}m^{-1}\), \(\beta_{5} = +3.15\cdot 10^{-11} ps^{5}m^{-1}\), \(\beta_{6} = - 8.00\cdot 10^{-14} ps^{6}m^{-1}\), and \(\beta_{7} = +2.50\cdot 10^{-16} ps^{7}m^{-1}\).

The right-side of Eq.(1) describes the NL effects, where \(\gamma_{0} = \frac{\omega_{0}n_{2}(\omega_{0})}{cA_{eff}} \) is the usual NL waveguide coefficient and the \(A_{eff}\) is the effective modal area. The value of \(\gamma_{0}\) is taken to be \(+0.0025 W^{-1} m^{-1}\), assuming that for fused silica at 800 nm we have \(n_{2}=+2.5 \cdot 10^{-20} m^{2}W^{-1}\), and the waveguide diameter is 10 \(\mu m\). We set the effective NL parameter for the propagation in the first segment as  $\gamma_{eff}= \gamma_{0} > 0$, and $\gamma_{eff}= \frac{\omega_{0}n_{2,eff}(\omega_{0})}{cA_{eff}} = -\gamma_{0}$ in the second segment. The respective negative NL effective refractive index, $n_{2,eff}$, can be provided by doping fused silica with silver NPs (with the volume fraction $\approx 10^{-4}$), as enabled by the Maxwell-Garnet theory \cite{zhavoronkov2011observation}. Another possibility to implement a negative NL parameter is offered by a BBO crystal for type-I phase-matching with \(\theta \approx 27.5^{\circ}\), where \(\theta\) is the angle between the input beam and optical c-axis of the BBO crystal \cite{vsuminas2017second}. 

The temporal derivative on the right-hand side of Eq.1 is associated with the third-order NL effects, such as self-steepening and optical-shock formation.

Concerning the integral term on the right-side of Eq.(1), the contribution of the delayed Raman response to the NL polarization is represented by \(f_{r}\) (equal to 0.18 for silica fibers \cite{Agrawal2013nonlinear}), which leads to effects such as the intrapulse Raman scattering and SSFS, while the local term represents the self-phase modulation (SPM) produced by the instantaneous electronic Raman contribution. Therefore, in this work, like in Ref.\cite{bose2016study}, we assume that both the self-focusing and defocusing waveguide segments are characterized by equal Raman terms.

\section{\label{sec:level3}Generation of soliton pairs by Multiple Temporal Compression (MTC)}

\subsection{\label{sec:level3.1}Pulse propagation in uniform media }

To examine the generation and evolution of bright solitons due to the pulse propagation in the composite medium with \(n_{2}\) of opposite signs, it is instructive to analyze the soliton dynamics considering only negative (Fig. \ref{fig:1}(a)-(c)) or positive (Fig. \ref{fig:1}(d,e)) nonlinearity. Then, we considered the propagation of a Gaussian input pulse \(A (0,T)=\sqrt{P_{0}}exp⁡\left(-0.5\left(\frac{1.665T}{T_{FWHM}}\right)^{2}\right)\), with central wavelength ($\lambda_{0} = 800 nm$), time duration ($T_{FWHM} = 90 fs$), and peak power ($P_{0} = 85 kW$). In the sample segment with $n_{2} < 0$ ($\gamma_{eff} = - 0.0025 W^{-1} m^{-1}$) with total length \(L = 150 mm\), the interplay between \(n_{2} < 0\) and the normal dispersion gives rise to the typical soliton fission around the central region of the pulse after the propagation distance \(L \approx 22 mm\). A modest spectral broadening driven by SPM due to the temporal compression of the pulse is shown in Fig. \ref{fig:1}(a)-(c). In cases of stronger spectral broadening it is possible to generate supercontinuum spectra throughout the visible to near-IR range, as the result of the first soliton fission \cite{dudley2002numerical}. 

As soon as the bright soliton acquires its form, we observe its robust evolution with high peak power \((\approx 3.8P_{0})\) under the influence of the intrapulse Raman scattering, which shifts the central wavelength to the red (Stokes) side during the propagation (Fig. \ref{fig:1}(c)). Once the bright soliton propagates under the action of the normal dispersion, its group velocity increases as its spectral center shifts towards longer wavelengths, leading to the soliton acceleration. This phenomenon exemplifies the SSFS effect and may strongly influence the temporal and spectral evolution of ultrashort solitons \cite{dudley2006supercontinuum, dudley2002numerical, demircan2008effects, bose2015experimental, braud2016solitonization, gordon1986theory, mitschke1986discovery, Agrawal2013nonlinear}.

Concerning the pulse propagation in the medium with the positive nonlinearity (\(\gamma_{0} = + 0.0025 W^{-1} m^{-1}\)), Figs. \ref{fig:1}(d,e) show the respective temporal and spectral evolution. The most important feature observed in this configuration is the temporal broadening which leads to reduction of the pulse peak power during the propagation. 

In the next section we explore the pulse propagation in the composite sample, with the first and second waveguide segments having \(n_{2} > 0\) and \(n_{2} < 0\), respectively. Figures \ref{fig:1}(d,e) are helpful for selecting the length of the first segment so as to avoid spectral saturation due to the decrease of the pulse peak power. Thus, the criterion for choosing the length of the first segment with positive \(n_{2}\) should be based on the balance between the NL phase accumulated from the action of the temporal SPM and temporal broadening of the pulse under the action of the normal dispersion. 

\subsection{\label{sec:level3.2}Pulse propagation in the composite medium }

In this subsection we investigate the pulse propagation initiated by the same Gaussian pulse considered in the previous section (\(\lambda_{0} = 800 nm, T_{FWHM} = 90 fs\), and \(P_{0} = 85 kW\)). The pulse propagates first in the segment with \(n_{2} > 0\), and then in the second segment with \(n_{2} < 0\). The calculation was performed for three different lengths of the first segment (\(L_{1} = 5, 10,\) and \(30 mm\)). 

 Figure 2(a-f) shows the spectral and temporal evolution of the pulse for each value of $L_{1}$. Notice that for $L_{1} = 5 mm$, besides the generation of a soliton pair caused by the double temporal compression, almost symmetrically with respect to the pulse’ center, the pair of bright solitons collide (at $L \approx 50 mm$) and fuse into one after propagation for few millimeters. Then, the fused soliton propagates with a higher peak power ,\(\approx 4P_{0}\), similar to the case displayed above in Fig. \ref{fig:1}(b).

\begin{figure*}
    \centering
    \begin{subfigure}[b]{0.312\textwidth}
     \centering
        \includegraphics[width=\textwidth]{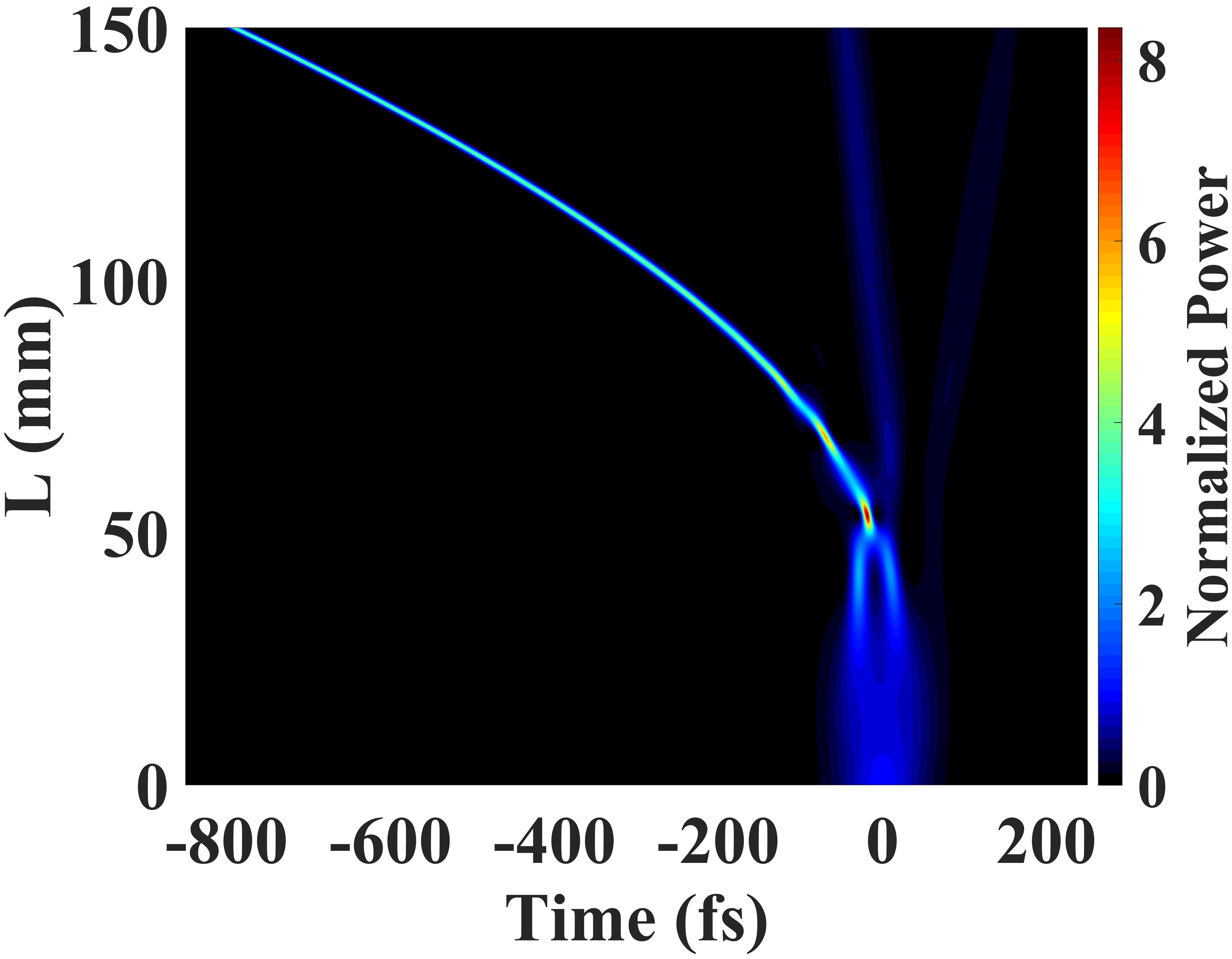}
         \caption{}
        \label{fig:2a}
    \end{subfigure}
    ~ 
    \begin{subfigure}[b]{0.32\textwidth}
     \centering
        \includegraphics[width=\textwidth]{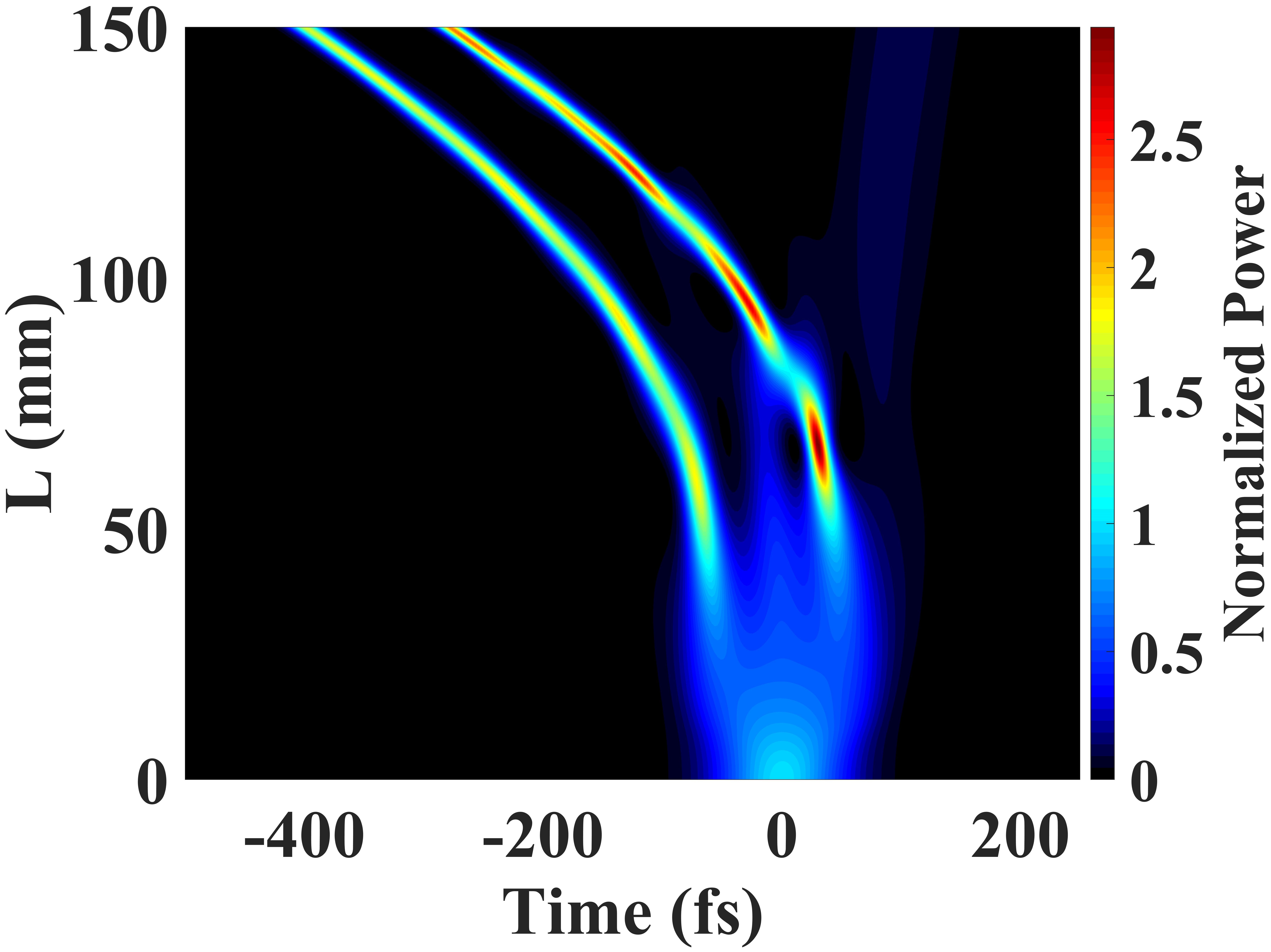}
         \caption{}
        \label{fig:2b}
    \end{subfigure}
    ~ 
    \begin{subfigure}[b]{0.32\textwidth}
     \centering
        \includegraphics[width=\textwidth]{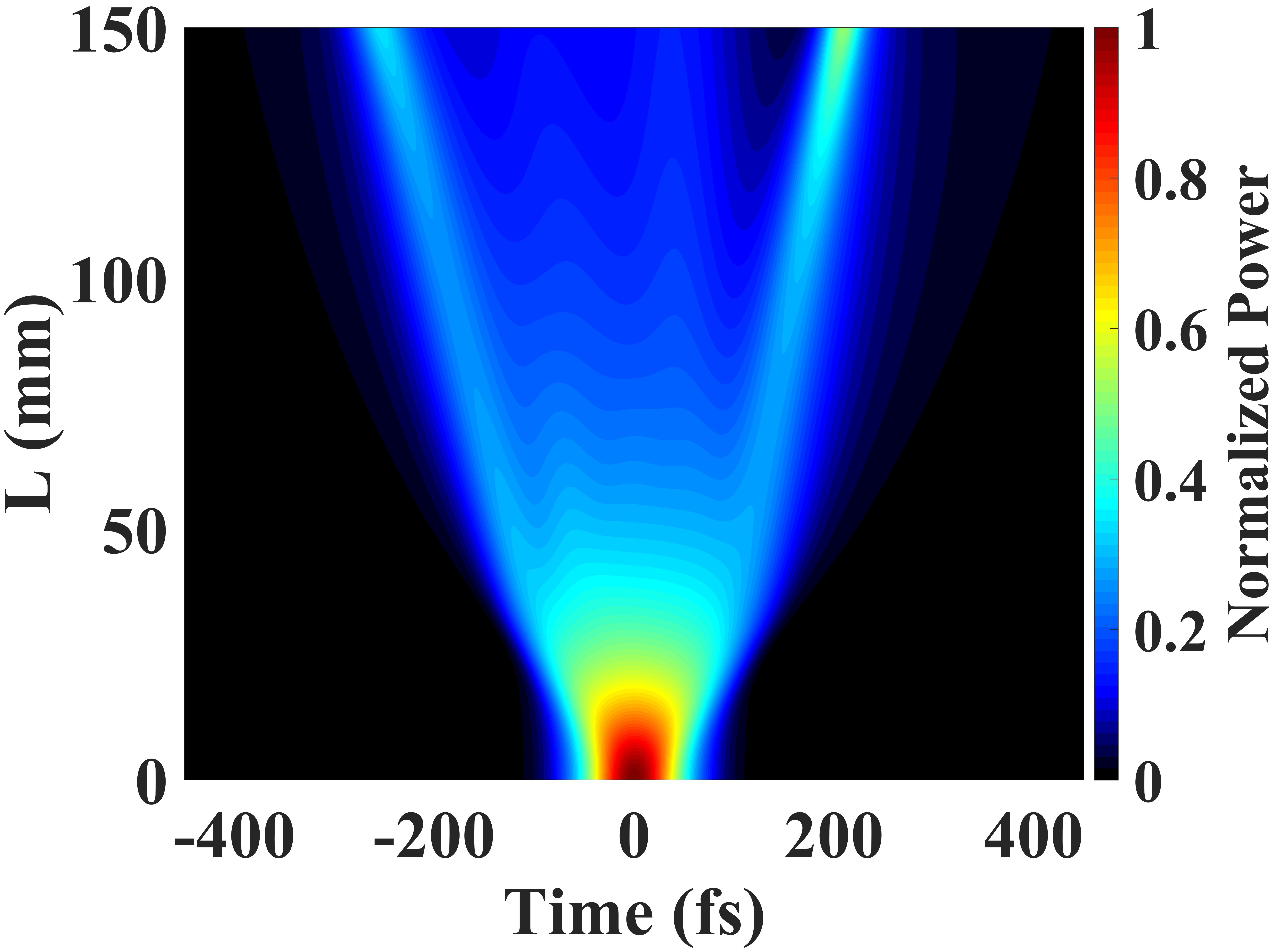}
         \caption{}
        \label{fig:2c}
    \end{subfigure}
    \begin{subfigure}[b]{0.32\textwidth}
     \centering
        \includegraphics[width=\textwidth]{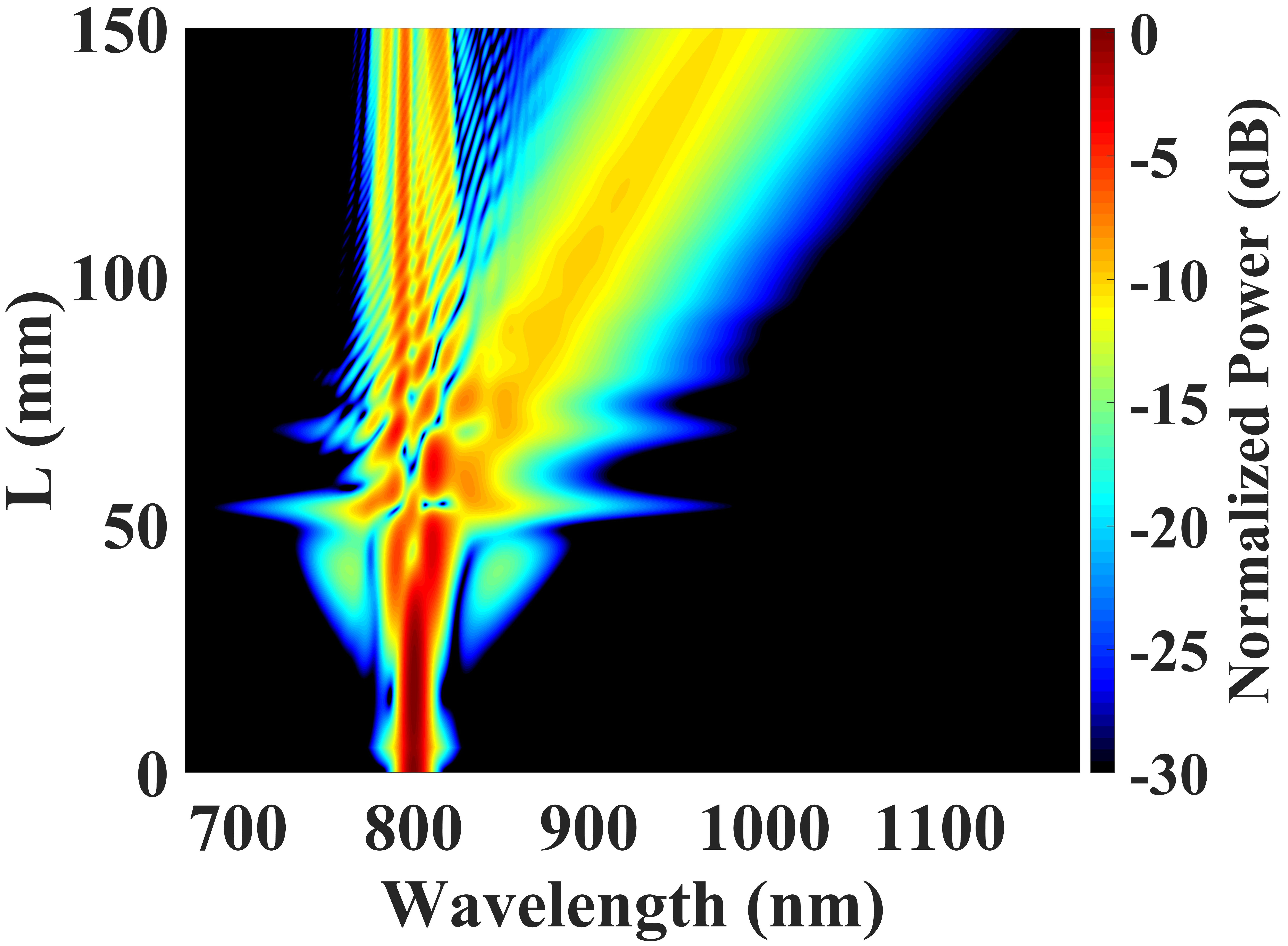}
         \caption{}
        \label{fig:2d}
    \end{subfigure}
  \begin{subfigure}[b]{0.32\textwidth}
        \includegraphics[width=\textwidth]{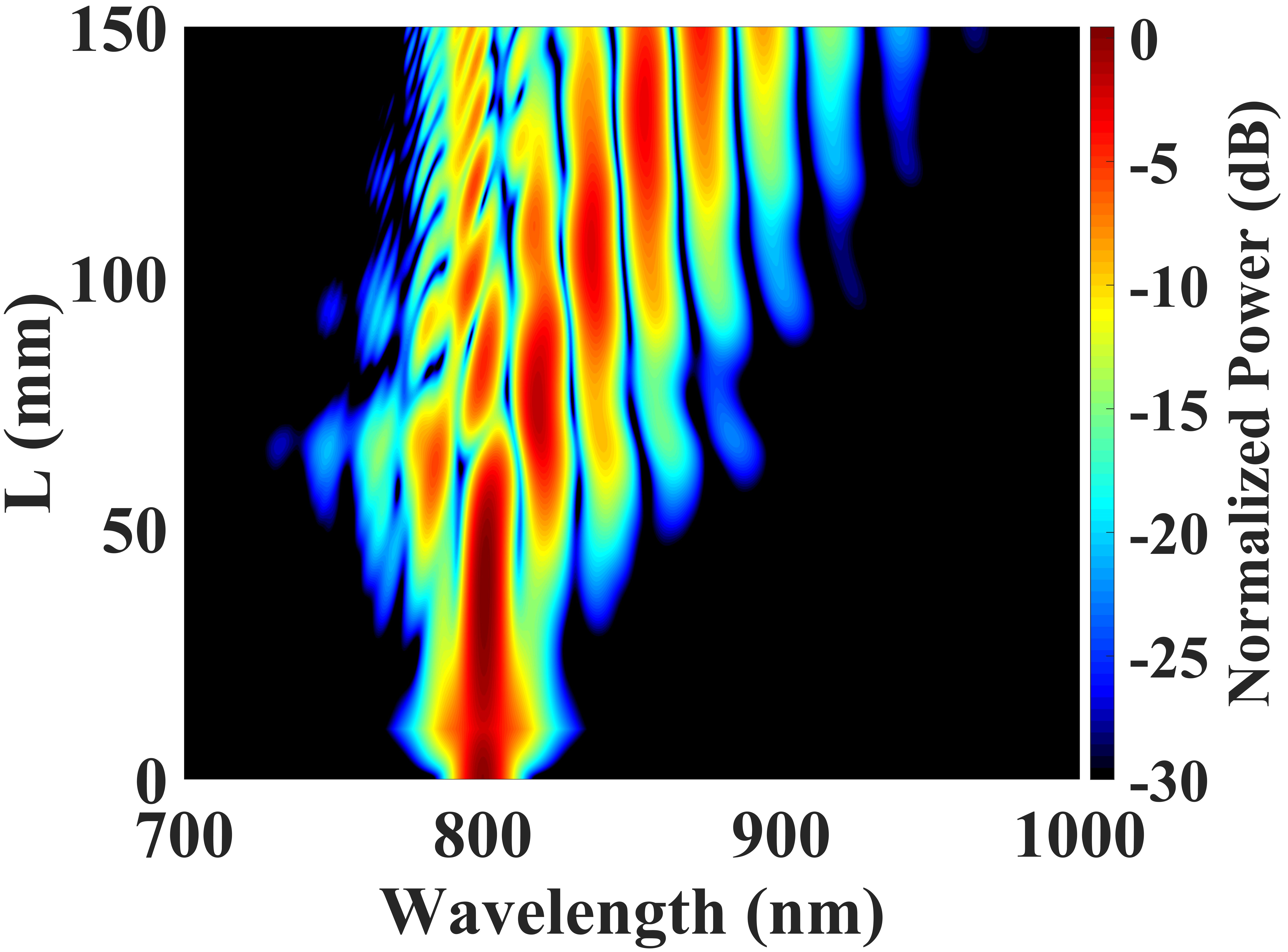}
         \caption{}
        \label{fig:2e}
    \end{subfigure}
\begin{subfigure}[b]{0.32\textwidth}
        \includegraphics[width=\textwidth]{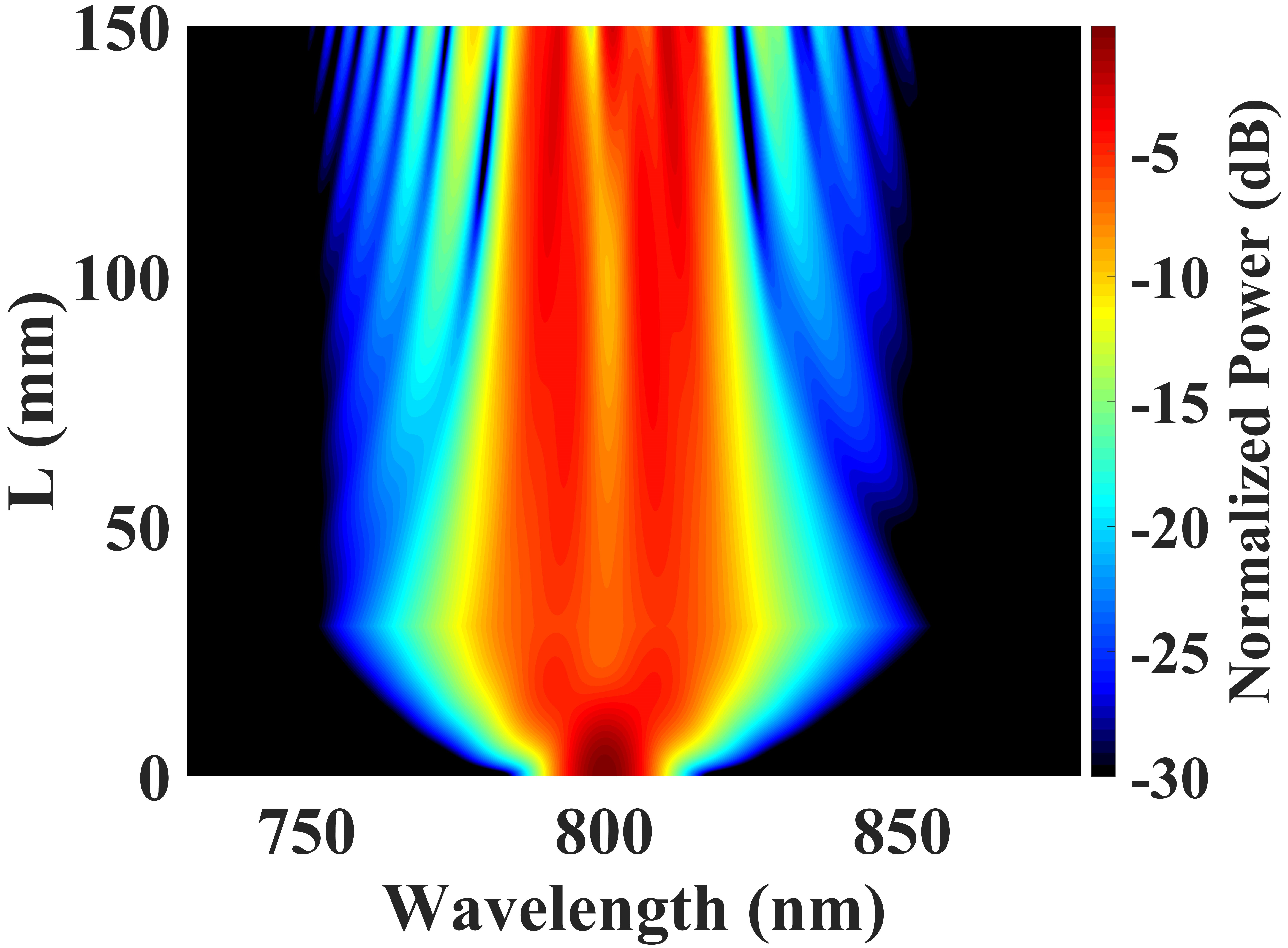}
         \caption{}
        \label{fig:2f}
    \end{subfigure}
    
    \caption{The generation of solitons via the MTC process: (a,b,c) The temporal evolution for the first segment with lengths \(L_{1}\) = 5, 10, and 30 mm, respectively. (d,e,f) The spectral evolution for the same cases.}
    \label{fig:2}
\end{figure*}

Because there is no soliton collision for \(L_{1} = 10 mm\) (Fig. \ref{fig:2}(b)), the soliton pair generated due to the double temporal compression gives rise to two fundamental bright solitons with similar peak powers, on both sides of the pulse. An essential feature of this dynamics is the contribution of the intrapulse Raman scattering, which is also similar for both solitons; they propagate close to each other as they experience similar red-shifts due to the SSFS. Therefore, increasing \(L_{1}\)  from 5 to 10 mm, it is worthy to mention that the scenarios explored in this work offer a potential for studying collisions between solitons with slightly different central wavelengths.

Note the oscillatory pattern in the spectral evolution displayed in Fig. \ref{fig:2}(e). These oscillations are attributed to the spectral superposition of the soliton pair, which is not observed in the case of \(L_{1} = 5 mm\)  because in such a case the soliton collision allows the soliton to experience relevant SSFS only at the earlier stage of the evolution, while later it propagates with a spectrum around 800 nm.

Concerning a longer first segment (\(L_{1} = 30 mm\)), as shown in Fig.\ref{fig:2}(c,f), even though the pulse has accumulated more NL phase in the first segment in comparison to the other cases, the temporal broadening of the Gaussian pulse in the normal-dispersion regime is enough to weaken the soliton generation in the second waveguide segment. This fact indicates the importance of knowing results of the propagation through the first segment, concerning the spectral saturation, to predict the soliton generation in the second medium.

\section{\label{sec:level4}Generation of solitons by Multiple Temporal Compression (MTC)}

In the previous sections instead of the most common scenario with the soliton generation occurring around the central region of the pulse, the soliton generation was observed from the double temporal-compression effect occurring on both sides of the pulse passing the composite medium. In this section, we explore the soliton generation by considering a Gaussian input pulse like that addressed in the previous section (\(\lambda_{0} = 800 nm, T_{FWHM} = 90 fs\)), but with higher peak powers.

First we fix the lengths of the first and second segments as \(L_{1} = 15 mm\) and \(L_{2} = 75 mm\). Then, like the previous section, the pulse propagates, at first, in the segment with \(\gamma_{0} = + 0.0025 W^{-1} m^{-1}\), and then in the second one with \(\gamma_{eff} = - 0.0025 W^{-1} m^{-1}\). In this configuration, the pulse demonstrates well-balanced temporal and spectral broadening in the first segment, which is required to observe the soliton generation in the second one.
\begin{table*}[]
\begin{tabular}{lllll}
\hline
\multicolumn{1}{|l|}{}    & \multicolumn{1}{l|}{\(P_{0} = 0.15 MW\)} & \multicolumn{1}{l|}{\(P_{0} = 0.30 MW\)} & \multicolumn{1}{l|}{\(P_{0} = 0.45 MW\)} & \multicolumn{1}{l|}{\(P_{0} = 0.60 MW\)} \\ \hline
\multicolumn{1}{|l|}{\(S_{LE}\)} & \multicolumn{1}{l|}{0.14 MW / 19 fs}   & \multicolumn{1}{l|}{0.32 MW / 12.5 fs}    & \multicolumn{1}{l|}{0.41 MW / 10 fs}    & \multicolumn{1}{l|}{0.50 MW / 9 fs}    \\ \hline
\multicolumn{1}{|l|}{\(S_{TE}\)} & \multicolumn{1}{l|}{0.16 MW / 18 fs}    & \multicolumn{1}{l|}{0.30 MW / 13 fs}    & \multicolumn{1}{l|}{0.33 MW / 12 fs}    & \multicolumn{1}{l|}{0.33 MW / 12 fs}    \\ \hline
                          &                           &                           &                           &                          
\end{tabular}
\caption{The peak power and time duration of the soliton leading edge (\(S_{LE}\)) and soliton trailing edge (\(S_{TE}\)) for different values of the input power. }
  \label{table:1}
\end{table*}

\begin{figure*}
    \centering
    \begin{subfigure}[b]{0.24\textwidth}
     \centering
        \includegraphics[width=\textwidth]{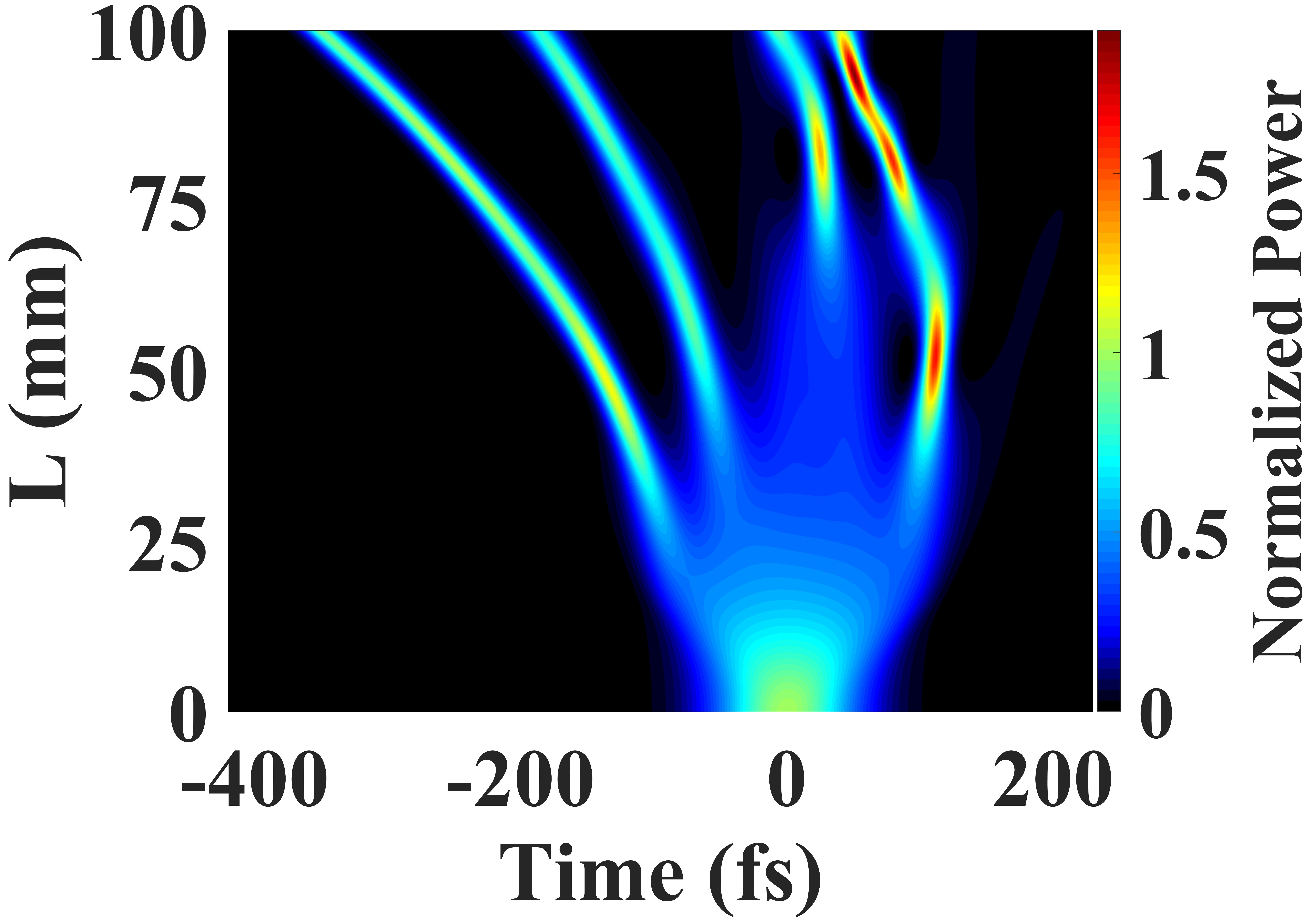}
         \caption{}
        \label{fig:3a}
    \end{subfigure}
    ~ 
    \begin{subfigure}[b]{0.24\textwidth}
     \centering
        \includegraphics[width=\textwidth]{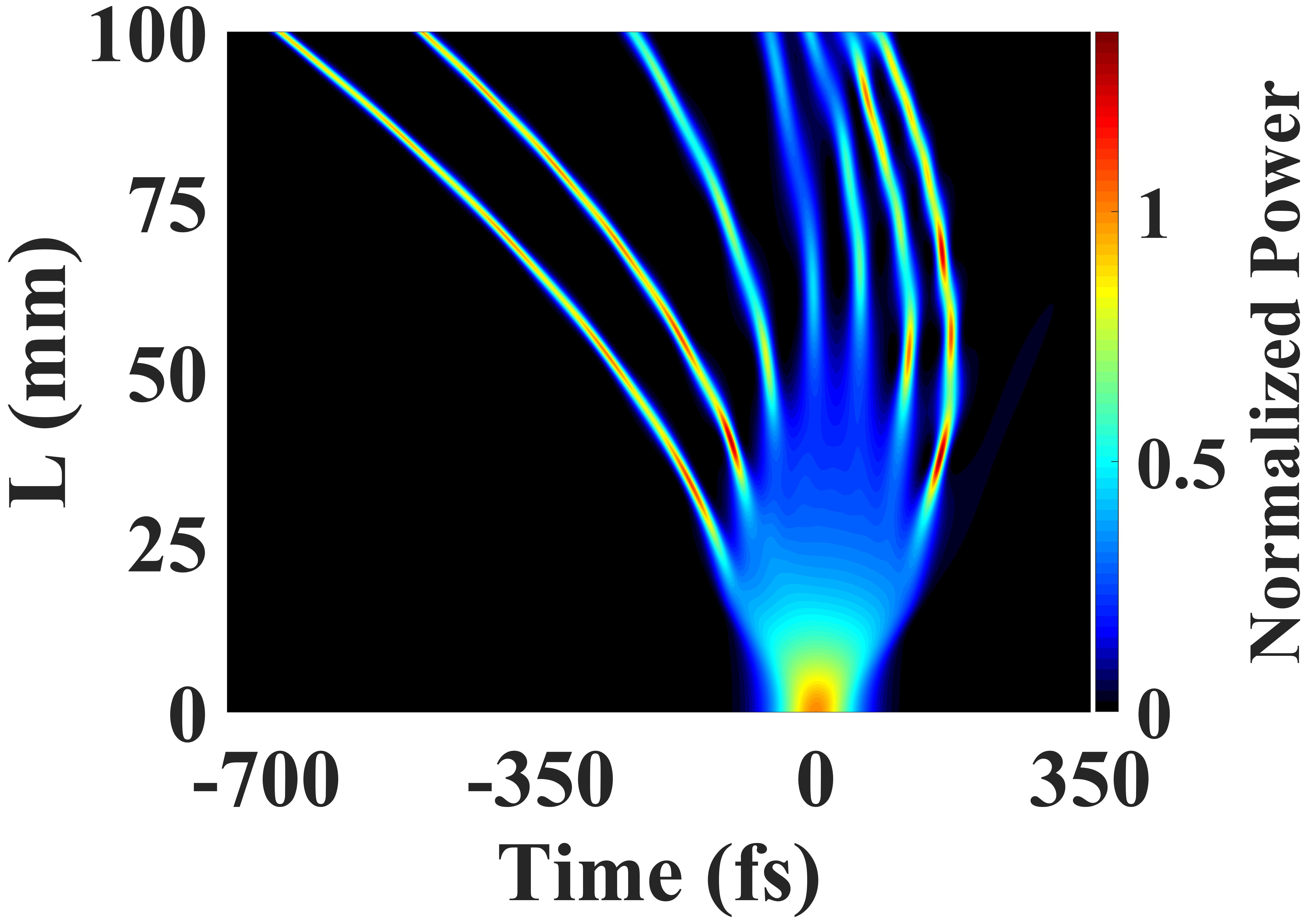}
         \caption{}
        \label{fig:3b}
    \end{subfigure}
    ~ 
    \begin{subfigure}[b]{0.242\textwidth}
     \centering
        \includegraphics[width=\textwidth]{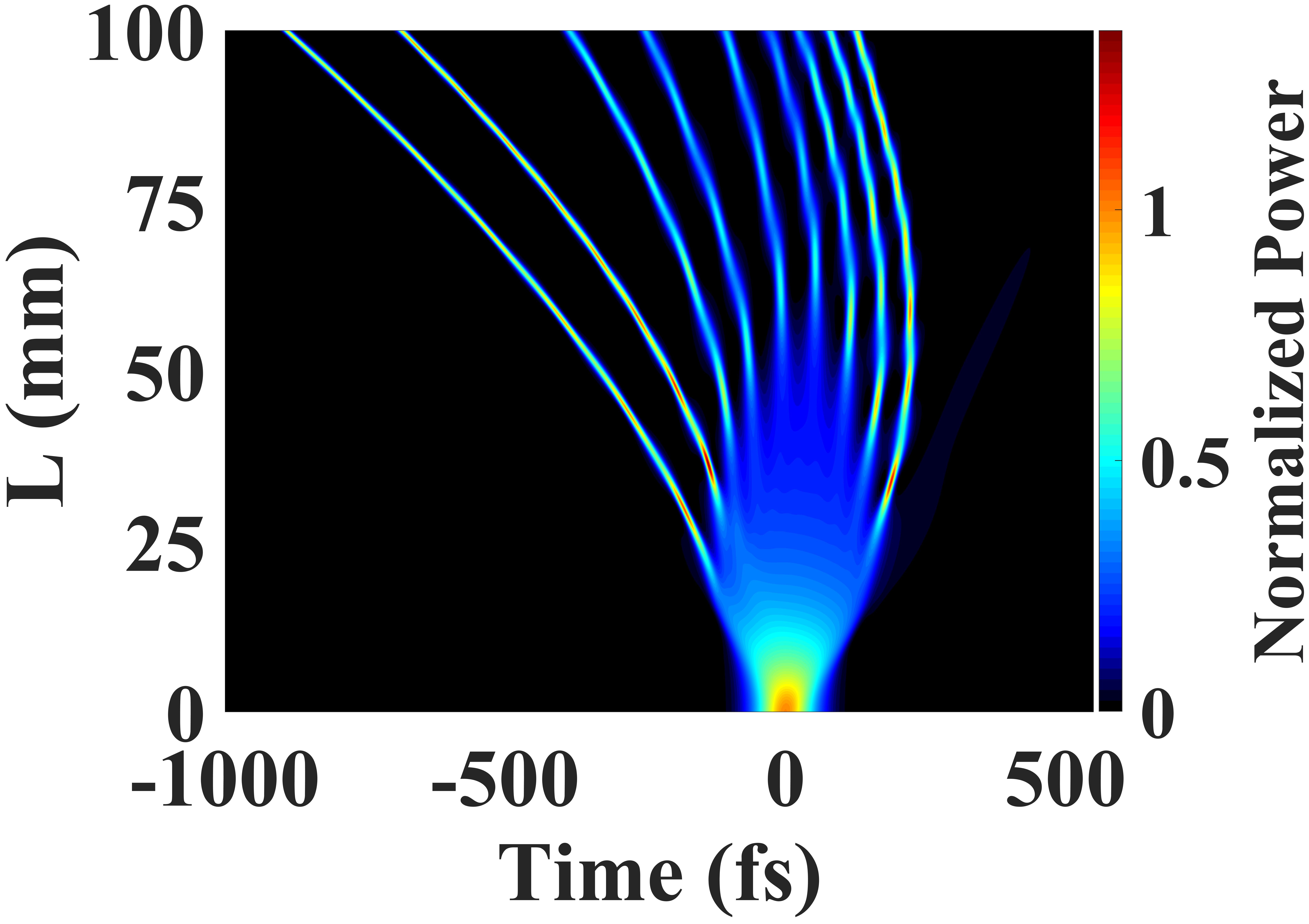}
         \caption{}
        \label{fig:3c}
    \end{subfigure}
    \begin{subfigure}[b]{0.24\textwidth}
     \centering
        \includegraphics[width=\textwidth]{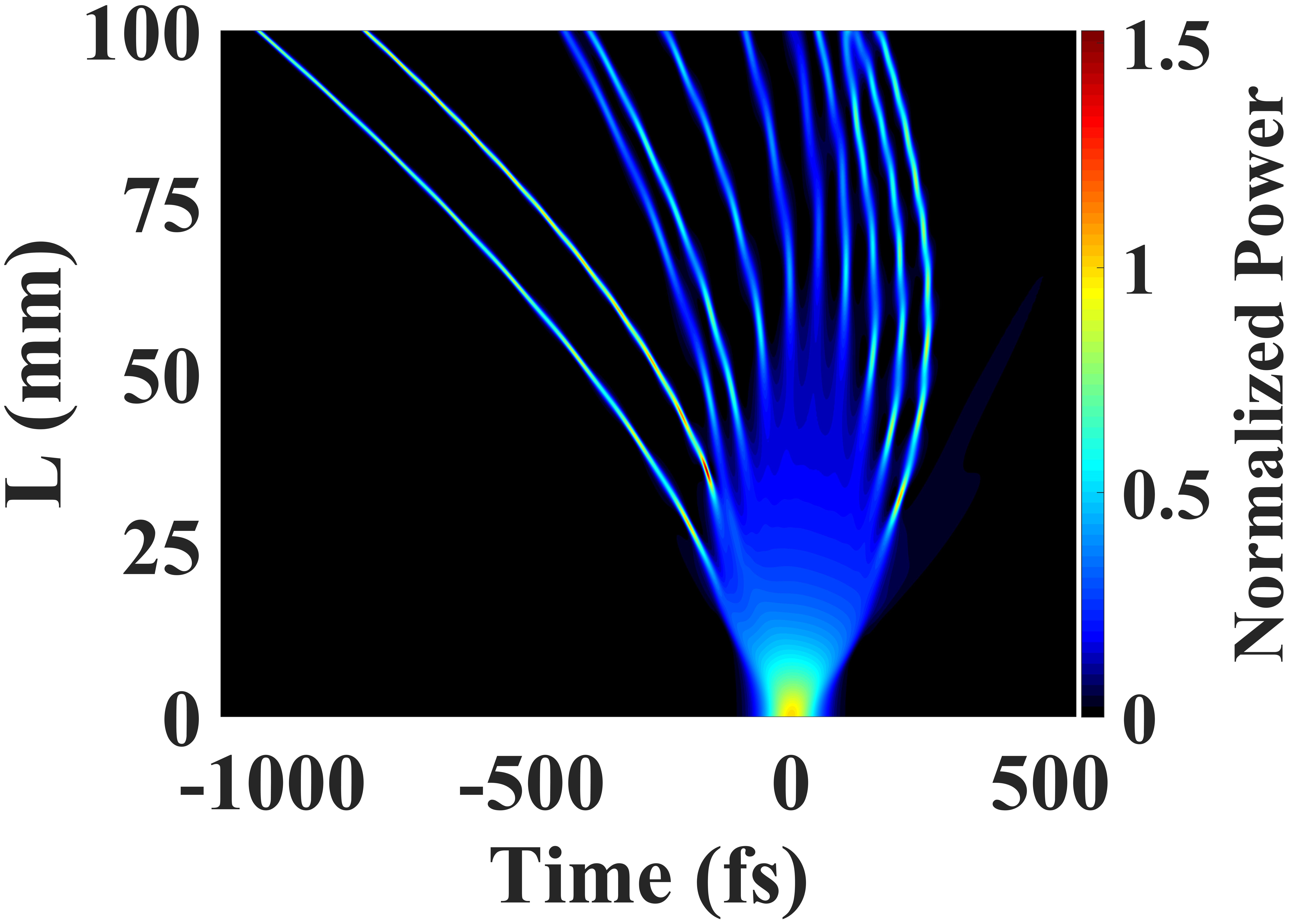}
         \caption{}
        \label{fig:3d}
    \end{subfigure}
  \begin{subfigure}[b]{0.236\textwidth}
        \includegraphics[width=\textwidth]{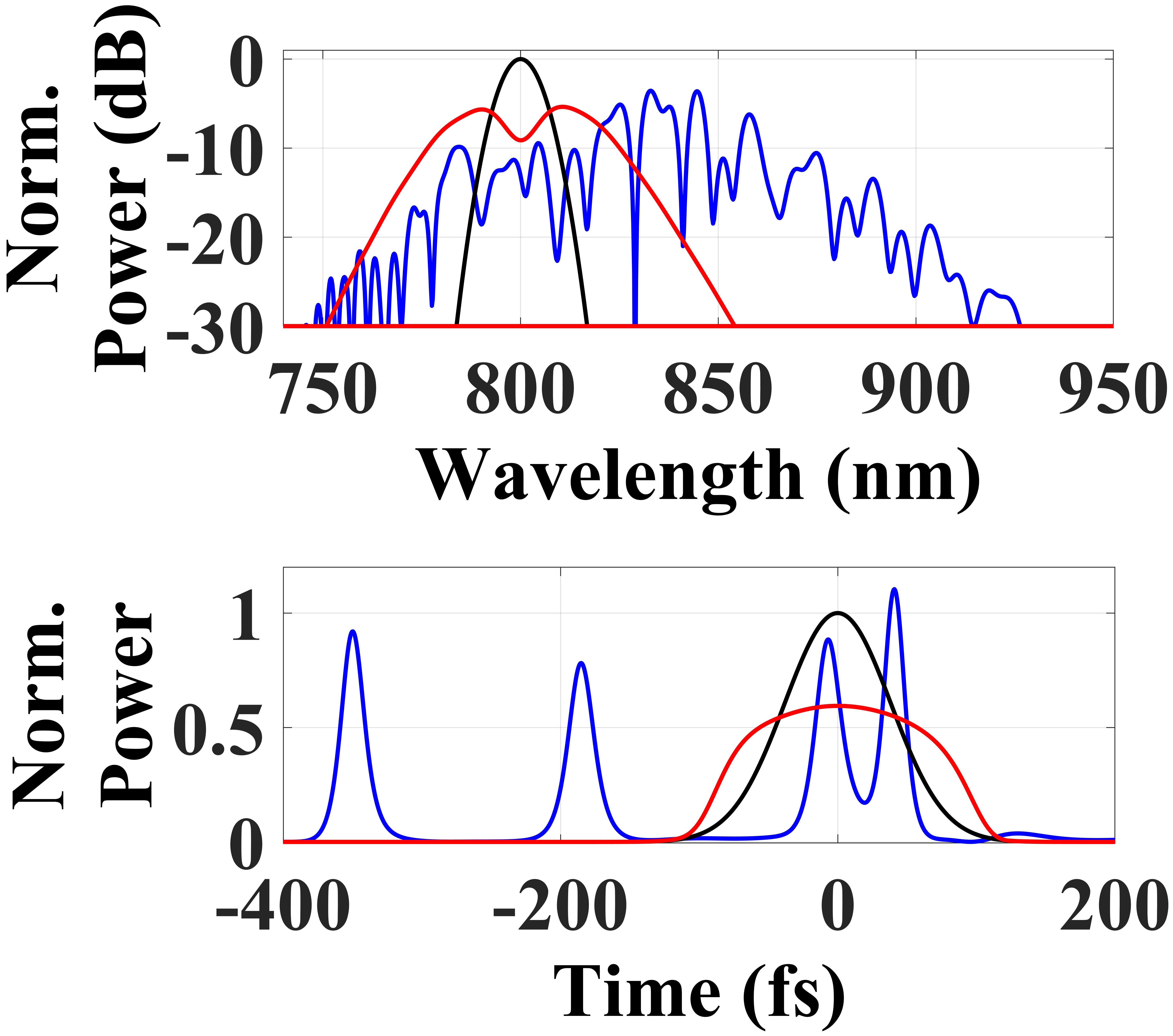}
         \caption{}
        \label{fig:3e}
    \end{subfigure}
\begin{subfigure}[b]{0.24\textwidth}
        \includegraphics[width=\textwidth]{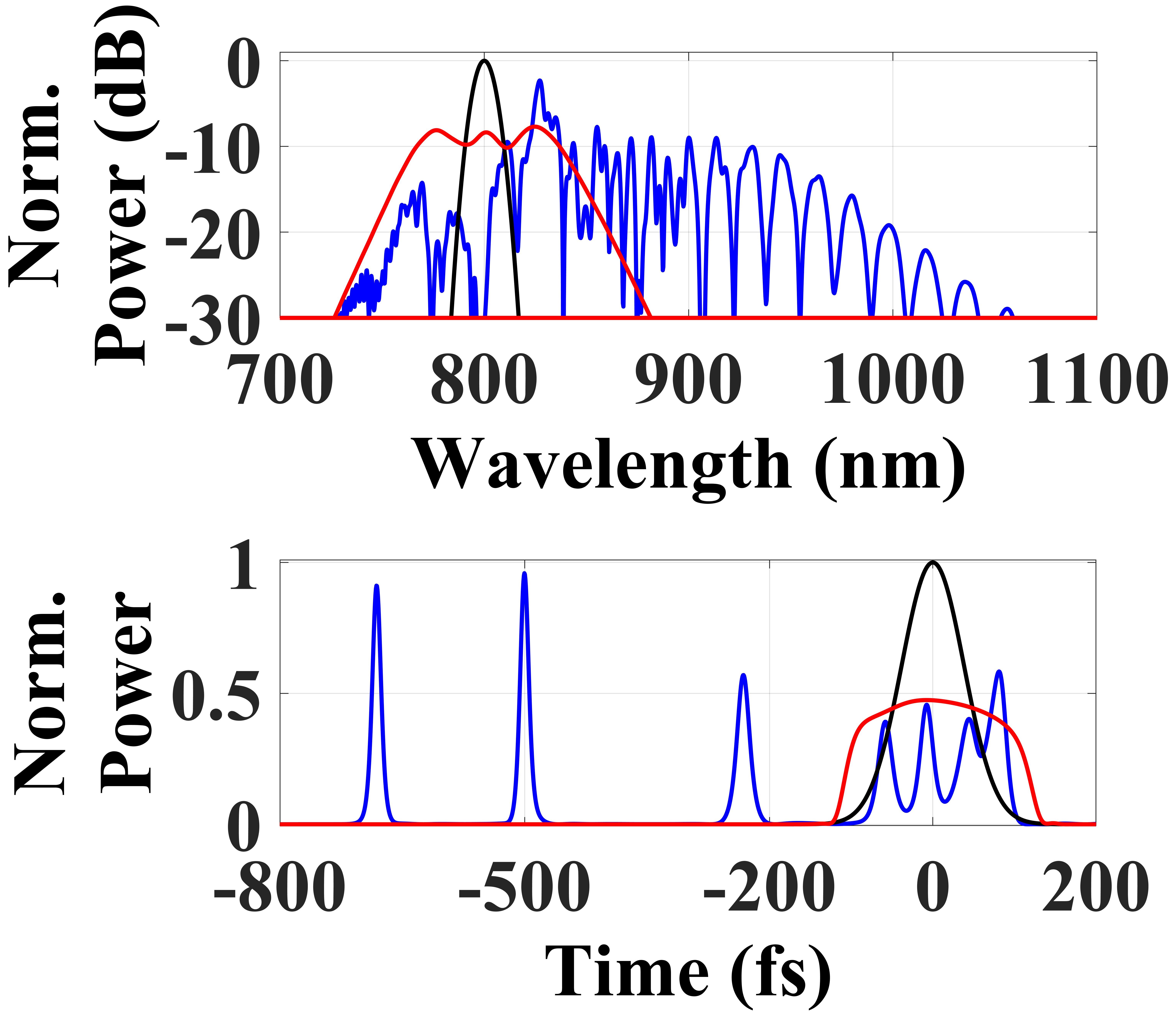}
         \caption{}
        \label{fig:3f}
    \end{subfigure}
    \begin{subfigure}[b]{0.235\textwidth}
        \includegraphics[width=\textwidth]{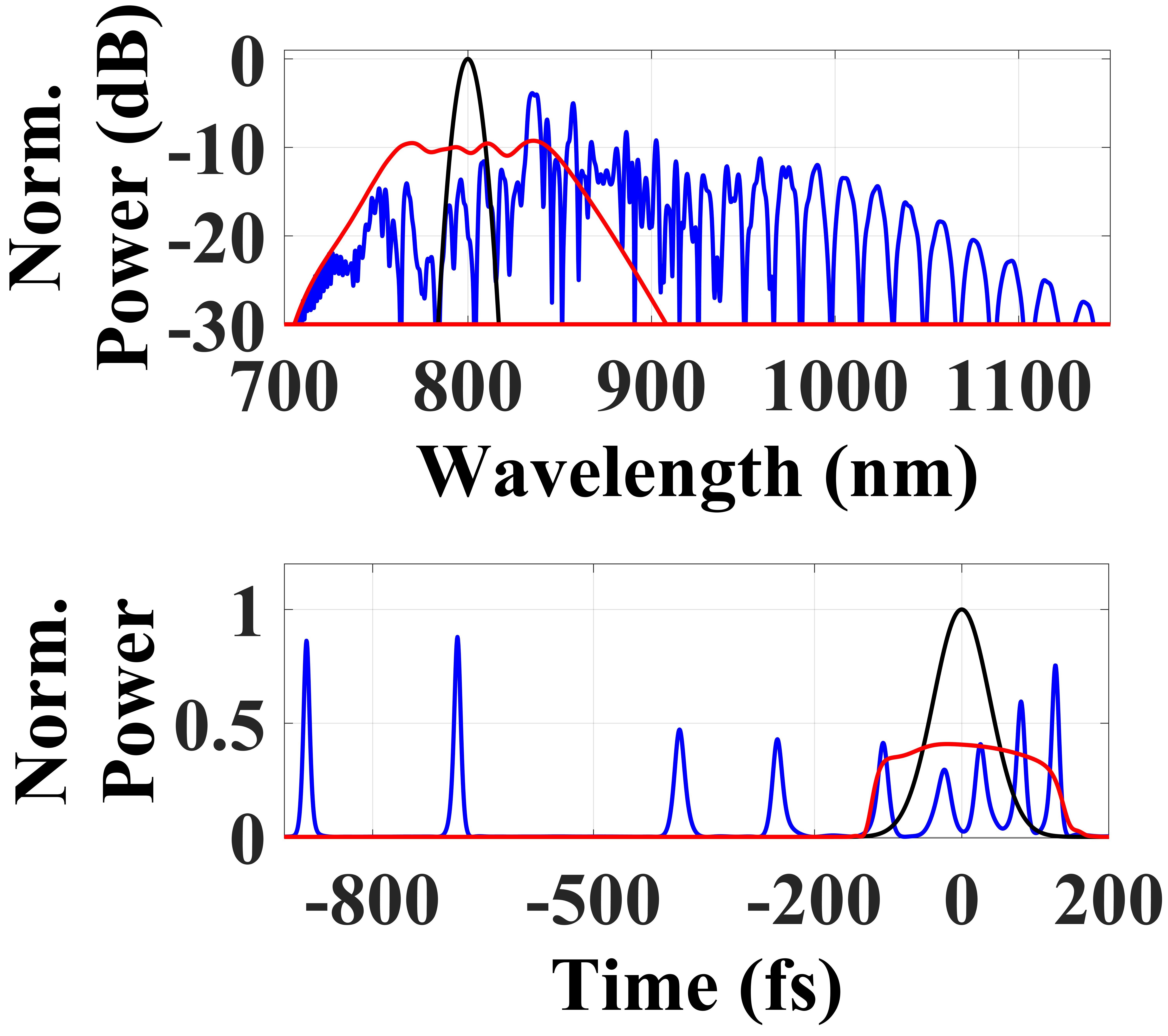}
         \caption{}
        \label{fig:3g}
        \end{subfigure}
        \begin{subfigure}[b]{0.233\textwidth}
        \includegraphics[width=\textwidth]{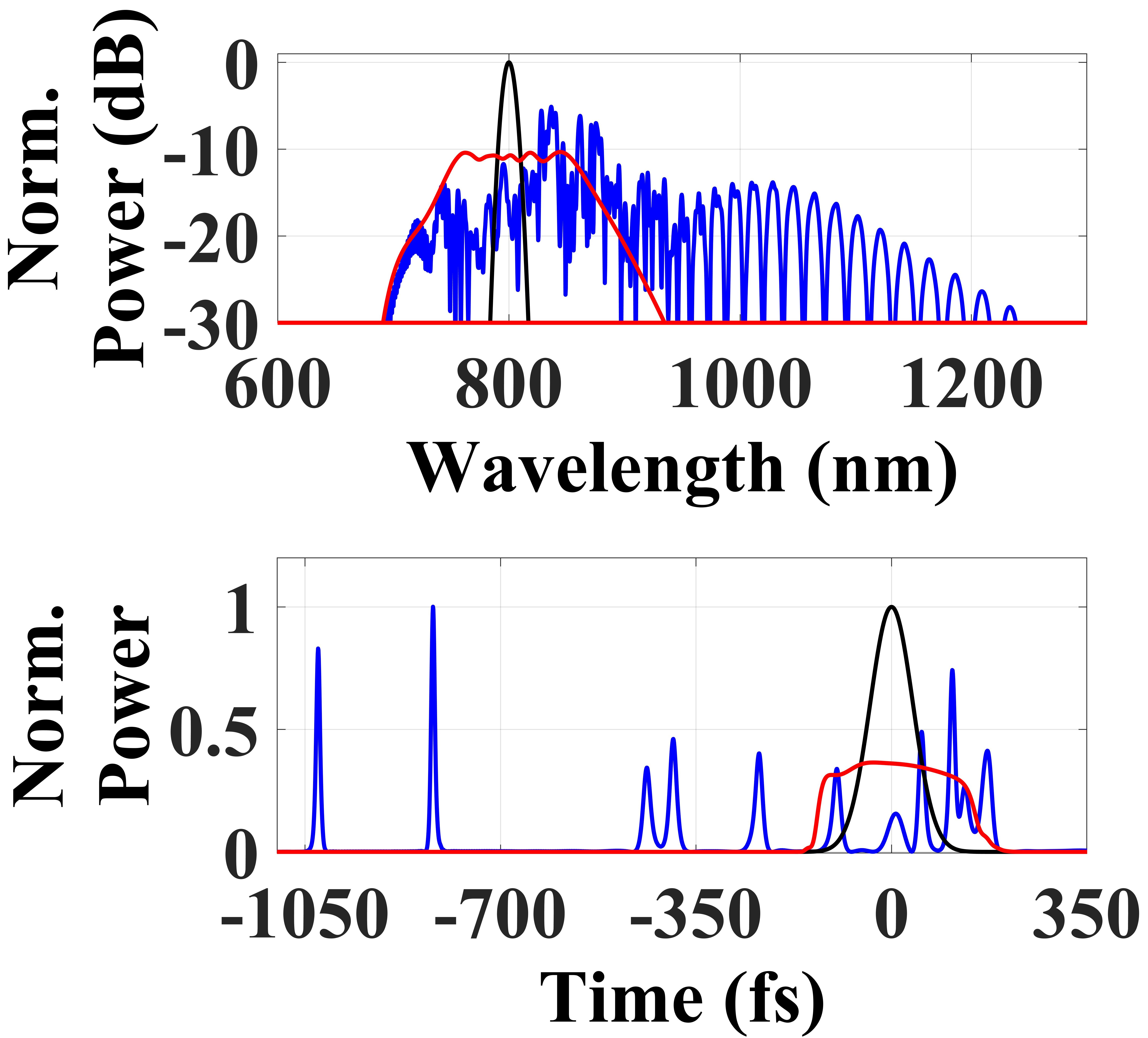}
         \caption{}
        \label{fig:3h}
           \end{subfigure}
    \caption{The generation of multiple solitons via the MTC process. (a-d) The temporal evolution from input pulses with peak powers equal to 0.15 MW, 0.30 MW, 0.45 MW, and 0.60 MW, respectively. (e-h) The temporal and spectral shapes during the pulse propagation: the black line illustrates the input pulse; the red line refers to the output of the SF segment for L = 15 mm; the blue line corresponds to the output of the second segment L = 100mm.}
    
    \label{fig:3}
\end{figure*}

The simulations were performed by taking the following values of input peak power: 0.15 MW, 0.30 MW, 0.45 MW, and 0.60 MW. For 0.15 MW a modest spectral broadening (from 750 nm to 870 nm) was observed in the first waveguide segment at -30 dB with respect to the maximum value. This power was enough to generate two pairs of bright solitons in the second segment, as shown in Fig. \ref{fig:3}(e) (the red curve). Furthermore, it is worthy to note that this configuration leads to the spectral shift towards the red side, chiefly due to the contribution of the intrapulse Raman scattering that acts individually onto each fundamental soliton according to the relation \(\Omega_{p} (z)\propto \frac{-z}{T_{FWHM}^{4} }\) \cite{Agrawal2013nonlinear}.

Considering input peak powers larger than 0.15 MW, more bright solitons are generated in the leading and trailing edges of the pulses due to the MTC, which is an intrinsic feature of the soliton generation driven by the pulse propagation in the segments with \(n_{2}\) of opposite signs and normal dispersion. Thus, the soliton generation starts from the first generation of the soliton pair, almost temporally symmetric with respect to the central region. The pair represents the soliton leading (\(S_{LE}\)) and trailing (\(S_{TE}\)) edges. As the pulse propagates in the second nonlinear segment additional fundamental soliton pairs are generated due to the MTC process. The generation of additional solitons which considers the soliton order of  \(N \approx 1\) due to the influence of the Raman term on the spectral and temporal evolution of the solitons, is illustrated by Eq.(2):  

\begin{equation}
    \begin{split}
N = \frac{T_{FWHM}}{1.665}\sqrt{\frac{\gamma_{eff}P_{0}^{soliton}}{\beta_{2}}} = \\ 
  = 0.1581\sqrt{P_{0}^{soliton}T_{FWHM}^{2}[ps]} \approx 1 .
  \label{eq:2}
 \end{split}
\end{equation}

To analyze the characteristics of the generated solitons, Table 1 shows typical values of peak power (\(P_{0}^{soliton}\)) and time duration (\(T_{FWHM}\)) of the \(S_{LE}\) and \(S_{TE}\)  according to the peak power of the Gaussian input pulse. Evaluating the increase in the input peak power from Table 1, temporal compression of \(S_{LE}\) was observed, accompanied by an increase in its peak power, while the \(S_{TE}\) did not show conspicuous variations in their peak powers and time durations for the Gaussian input pulse with \(P_{0} > 0.30 MW\). Concerning the secondary solitons generated in this case, we note that they tend to be slightly broader, as they are generated closer to the central region of the pulse, without a strong variation in their peak powers, as shown in Fig. \ref{fig:3}(e-h).

Another interesting feature revealed by the analysis is presented in Fig.4, which shows the relation between the input peak power of the Gaussian input pulse (the horizontal axis) and the respective number of generated solitons (black dots), following the relation between the input peak power of the Gaussian pulse and its equivalent soliton orders, \emph{viz}., \(N(P_{0}) \propto \sqrt{P_{0}}\) (the blue curve in Fig. 4). For instance, considering the Gaussian input pulse corresponding to an $N$-th order soliton, it creates approximately \(N\) fundamental solitons during the propagation, thus demonstrating a noteworthy phenomenon of the energy redistribution among the bright solitons.

\begin{figure}
    \centering
    \includegraphics [width=.40\textwidth]{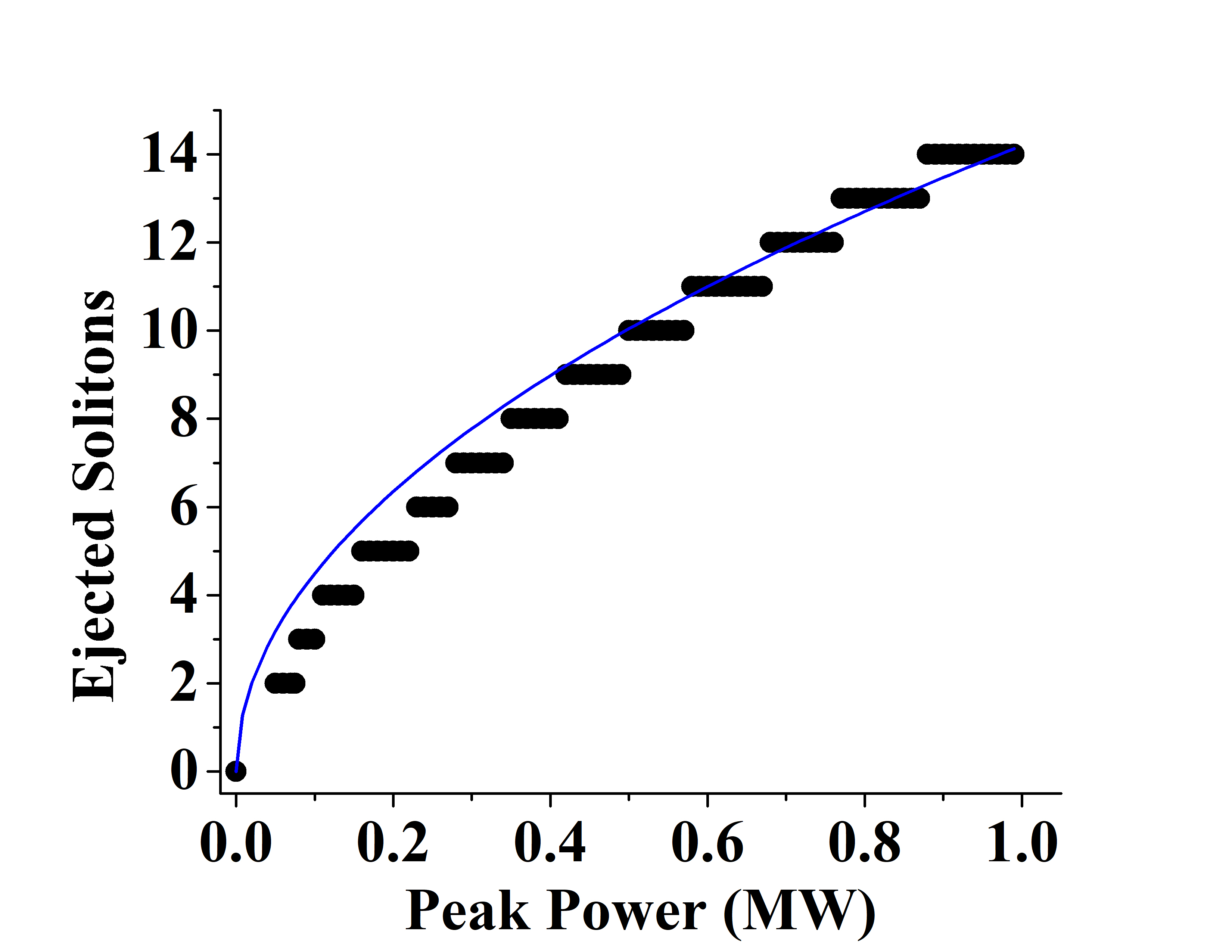}
    \caption{The number of ejected solitons by the MTC as a function of the input peak power (black segments). The equivalet soliton order (see the left side of Eq.(2)) of the input pulse taking into account its time duration and peak power (the blue curve).}
    \label{fig:ejected}
\end{figure}

\begin{figure}
    \centering
    \begin{subfigure}[b]{0.41\textwidth}
     \centering
        \includegraphics[width=\textwidth]{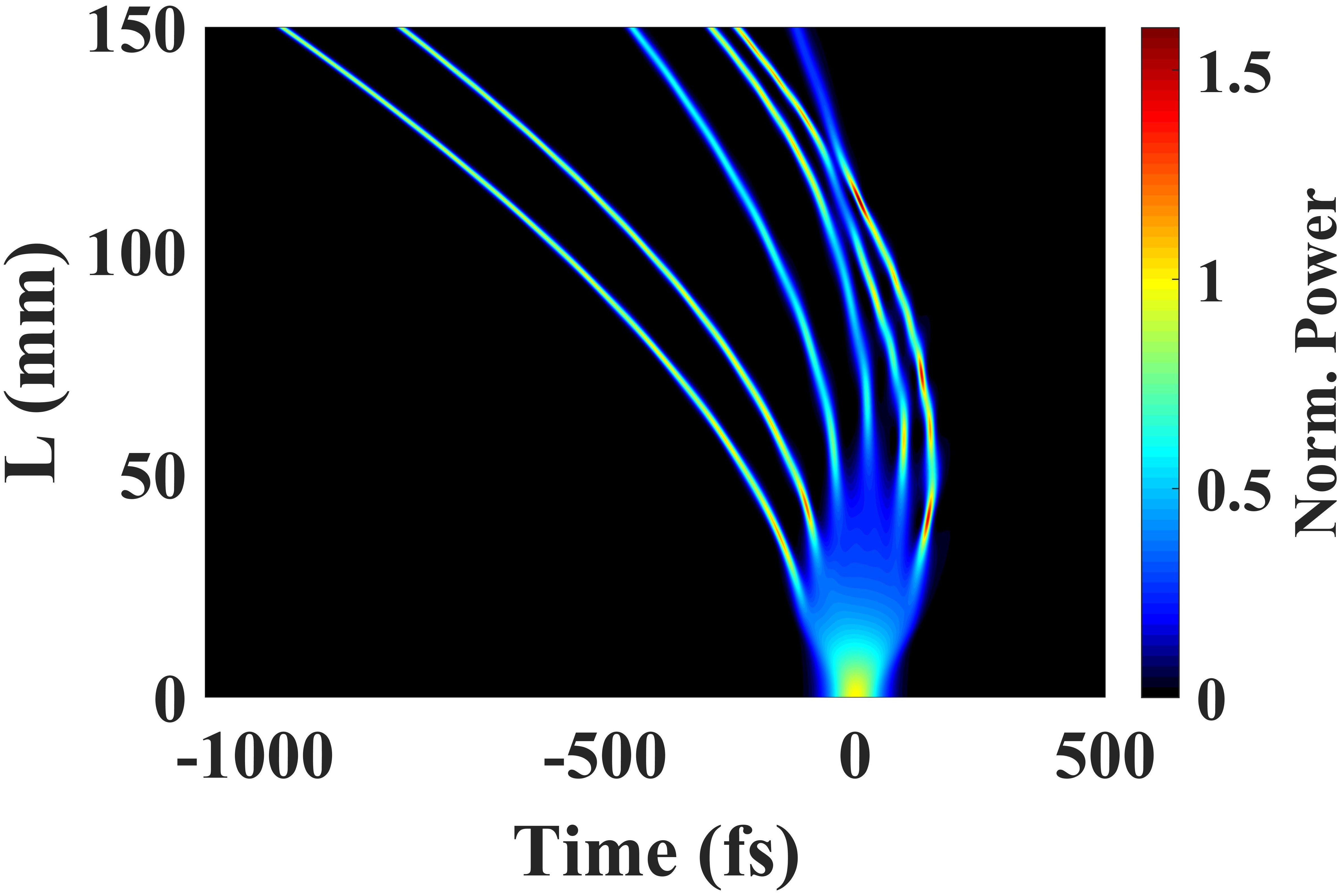}
         \caption{}
        \label{fig:5a}
    \end{subfigure}
    ~ 
    \begin{subfigure}[b]{0.41\textwidth}
     \centering
        \includegraphics[width=\textwidth]{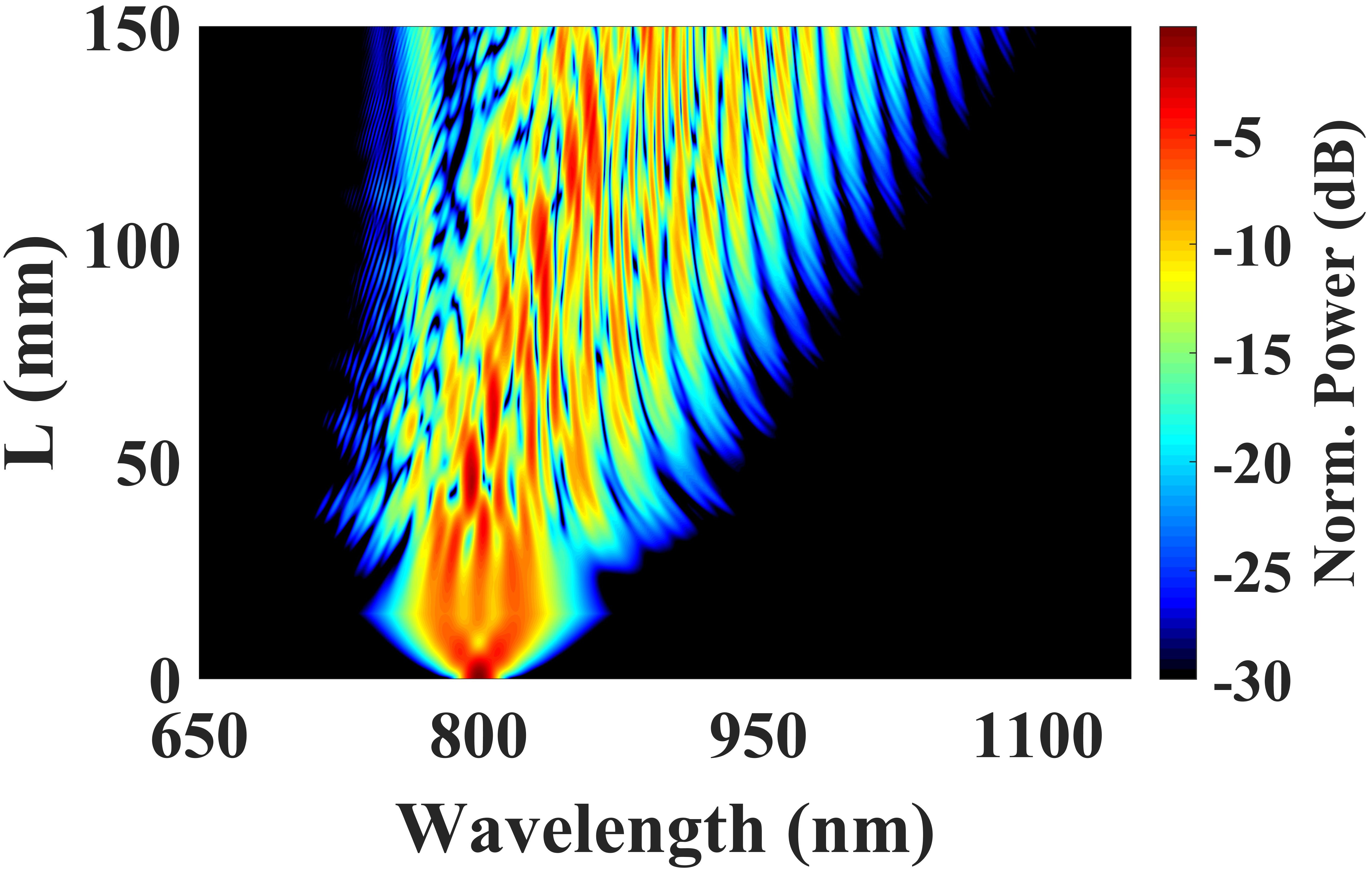}
         \caption{}
        \label{fig:5b}
    \end{subfigure}
    ~ 
    \begin{subfigure}[b]{0.41\textwidth}
     \centering
        \includegraphics[width=\textwidth]{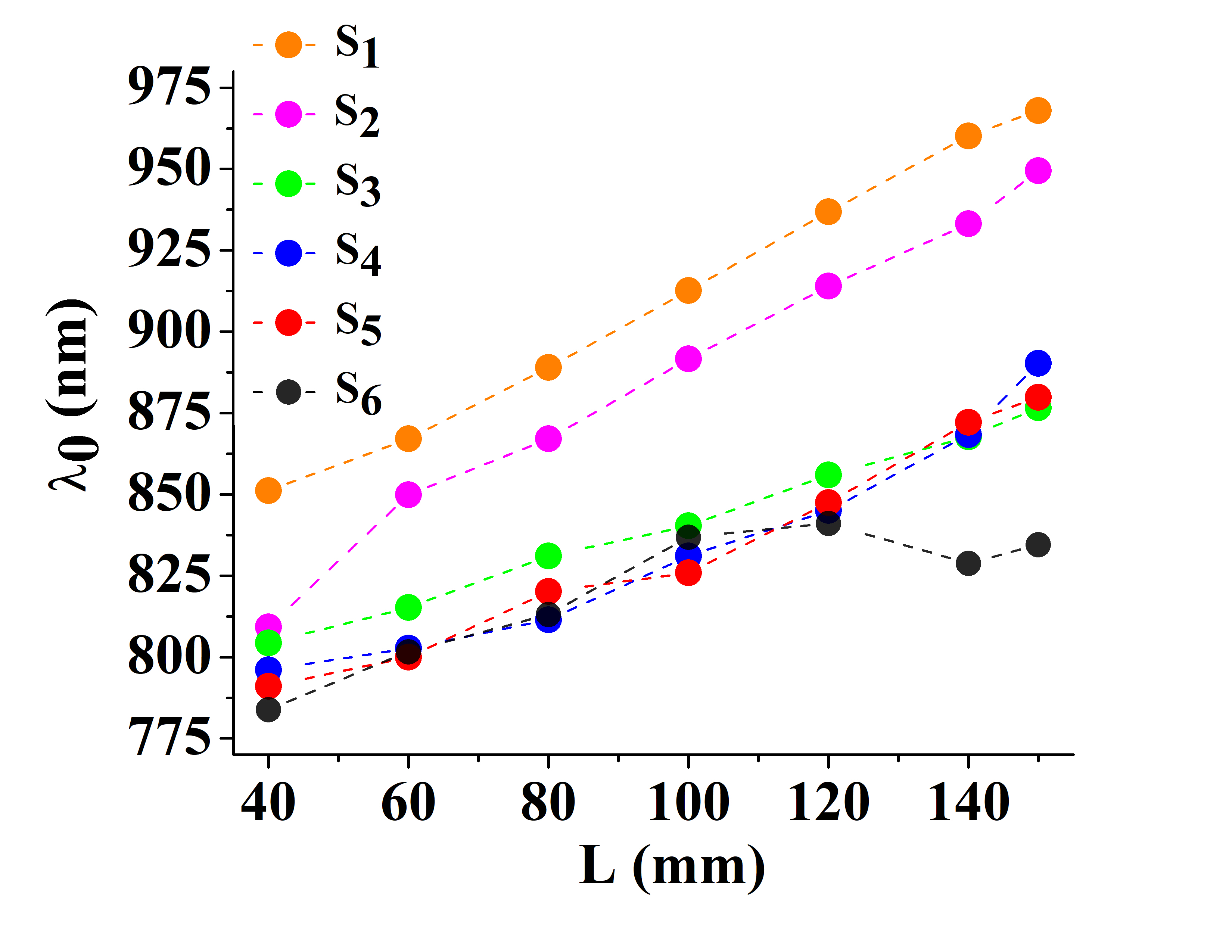}
         \caption{}
        \label{fig:5c}
    \end{subfigure}
    \caption{(a,b) The temporal and spectral evolution in a system with \(L_{1} = 15 mm\) initiated by the input pulse with peak power 0.25 MW. (c) The evolution of the central wavelength (\(\lambda_{0}\)) of the solitons \(S_{1}\), \(S_{2}\), \(S_{3}\), \(S_{4}\), \(S_{5}\), and \(S_{6}\) during the pulse propagation.}
    \label{fig:5}
\end{figure}

Because of the discrete nature of the soliton creation, there are different values of input peak power resulting in the same number of the generated solitons as illustrated by horizontal black segments in Fig. 4; in the degeneracy ranges, small temporal compressions of the solitons were observed with the increase of the input peak power. This feature is more relevant for higher input peak powers, making the degeneracy ranges broader.

Concerning the soliton dynamics illustrated by Fig. \ref{fig:3}, when SSFS pushes the central soliton wavelength towards longer values, conspicuous soliton acceleration occurs in the normal-dispersion regime, leading to increase of the soliton's group velocity as they individually experience the action of the SSFS effect. Then, to investigate the temporal and spectral evolution of the bright solitons in this setting, we focused on the specific case with the input peak power 0.25 MW, $L_{1} = 15 mm$, and second segment length $L_{2} = 135 mm$. 

In this way, Fig. \ref{fig:5}(a,b) shows the temporal and spectral evolution of the Gaussian pulse with the same parameters as in Fig. \ref{fig:ejected} (except for \(L_{2}\)). We applied numerical filters to temporally isolate each fundamental soliton during its propagation, and then applied the Fourier transform to produce their spectra. As shown in Fig. \ref{fig:5}(c), when all solitons have already been generated (at \(L = 40 mm\)), it is straightforward to conclude that the solitons generated at the frontal region of the pulse are red-shifted (\(\lambda_{0} > 800 nm\)), and ones generated at the back of the pulse are blue-shifted (\(\lambda_{0} < 800 nm\)), in such a way that the soliton's central wavelength slightly decreases while proceeding from the leading edge of the pulse (\(S_{1}\), \(S_{2}\), and \(S_{3}\)) to the trailing edge (\(S_{4}\), \(S_{5}\), and \(S_{6}\)).

Once the two solitons generated at the early stage of the evolution in the leading edge (\(S_{1}\) and \(S_{2}\)) are temporally isolated, there is no soliton collision which would change their central wavelengths, and hence their group velocities. Therefore, as both solitons have similar peak powers and pulse durations, they experience a similar SSFS, which provides a continuous soliton acceleration in the normal-dispersion regime. Concerning the solitons generated in the trailing edge, each blue-shifted soliton (\(S_{4}\), \(S_{5}\), and \(S_{6}\)) experience its own SSFS, a trend to acquire longer wavelengths is seen for all solitons in Fig. \ref{fig:5}(a),  from the observation of the increase in their group velocities.

In the case of the propagation length \(L = 60 mm\), due to the higher peak power of \(S_{6}\) in comparison to neighboring solitons (\(S_{4}\) and \(S_{5}\)), \(S_{6}\) undergoes stronger SSFS, so that the central wavelengths of \(S_{4}\), \(S_{5}\), and \(S_{6}\) coalesce to the value of 800 nm. Hence, close to \(L = 60 mm\) solitons \(S_{4}\), \(S_{5}\), and \(S_{6}\) have similar group velocities, propagating together with the central region of the pulse if the Raman contribution suddenly turns off.

Although strong SSFS suffered by the \(S_{6}\) soliton produces notable soliton acceleration, after propagation in a distance \(L \approx 100 mm\), its central wavelength exceeds the wavelengths of neighboring solitons. This trend stops when the \(S_{6}\) soliton transfers part of its energy to the \(S_{5}\) soliton through an inelastic collision at \(L \approx 120 mm\). After collision, the \(S_{5}\) soliton quickly shifts its central wavelength to longer values, while the \(S_{6}\) soliton experiences a transient effect due to the soliton depletion. Thus, only after the propagation distance of \(L = 140 mm\), the \(S_{6}\) soliton demonstrate conspicuous SSFS.

To summarize, the generation of multiple solitons through the MTC process is initiated by the pulse propagation in the first segment under the action of the normal dispersion and SF. The respective output signal carries positive chirp, as shown in the spectrogram at $L = L_{1} = 15 mm$ in Fig.6 (a). At this point, the pulse has accumulated positive nonlinear and linear phases. 

Because the second segment applies the normal dispersion and negative nonlinearity to the propagating pulse, there is a transient length associated with the NL phase compensation. It means that the frequency generation by SPM occurs at opposite edges with respect to the frequency generation in the first segment. For instance, the leading edge of the pulse supports the generation of red-shifted components in the first waveguide segment and blue-shifted ones in the second segment. Once the red-shifted components propagate faster than their blue-shifted counterparts (in the normal-dispersion regime), the generation of multiple solitons through the temporal compression after passing the transient length is enabled at both edges of the pulse. In Fig. 5(b), the transient length is between $L = 15 mm$ and $L = 30 mm$, where the accumulation of the positive linear phase prevents the full compensation of the pulse chirp, what would prevent the MTC process.

After the transient length, the first soliton pair is generated at both edges of the pulse, as shown in the spectrogram for $L = 30 mm$ in Fig.6 (b). Then, almost the whole pulse is split into multiple solitons. In particular, at $L = 75 mm$, the system creates six solitons (see Fig. 6 (d)), and the above-mentioned soliton collision between $S_5$ and $S_6$ at $L = 120 mm$ can be seen in the spectrogram in Figs. 6(e,f).

\begin{figure}
  
    \begin{subfigure}[b]{0.315\textwidth}

        \includegraphics[width=\textwidth]{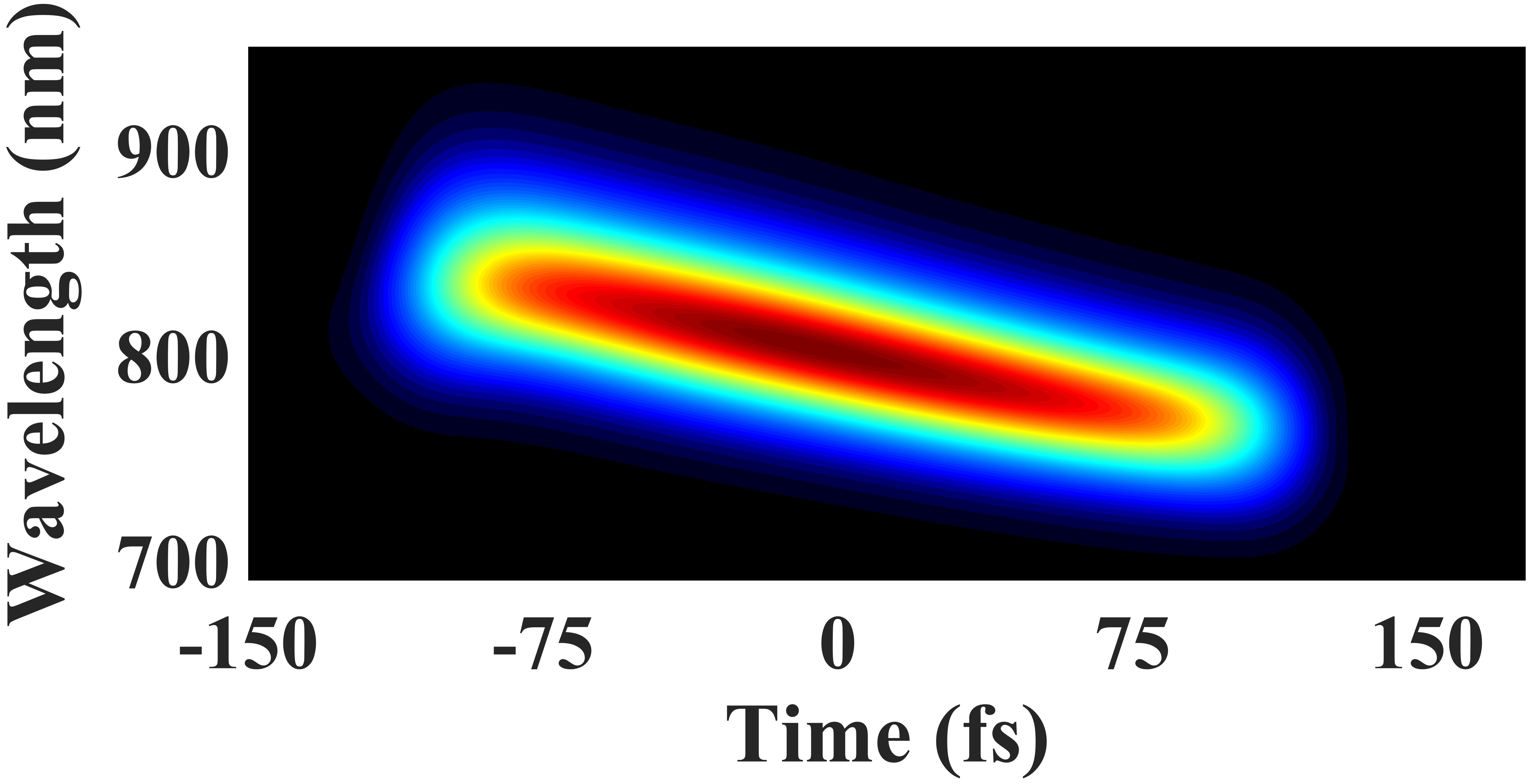}
         \caption{}
        \label{fig:6a}
    \end{subfigure}
   
    \begin{subfigure}[b]{0.315\textwidth}

        \includegraphics[width=\textwidth]{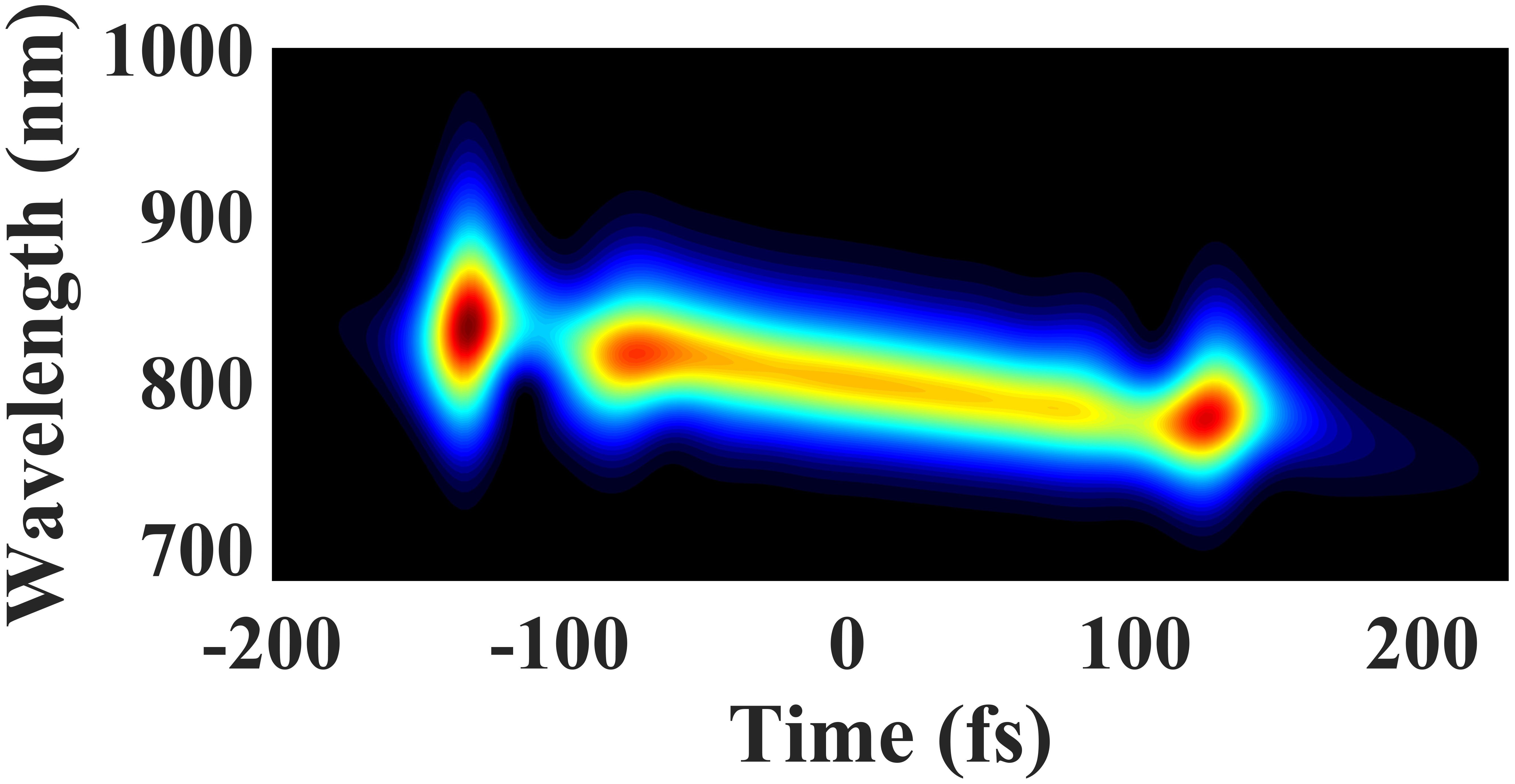}
         \caption{}
        \label{fig:6b}
    \end{subfigure}
    ~ 
    \begin{subfigure}[b]{0.315\textwidth}

        \includegraphics[width=\textwidth]{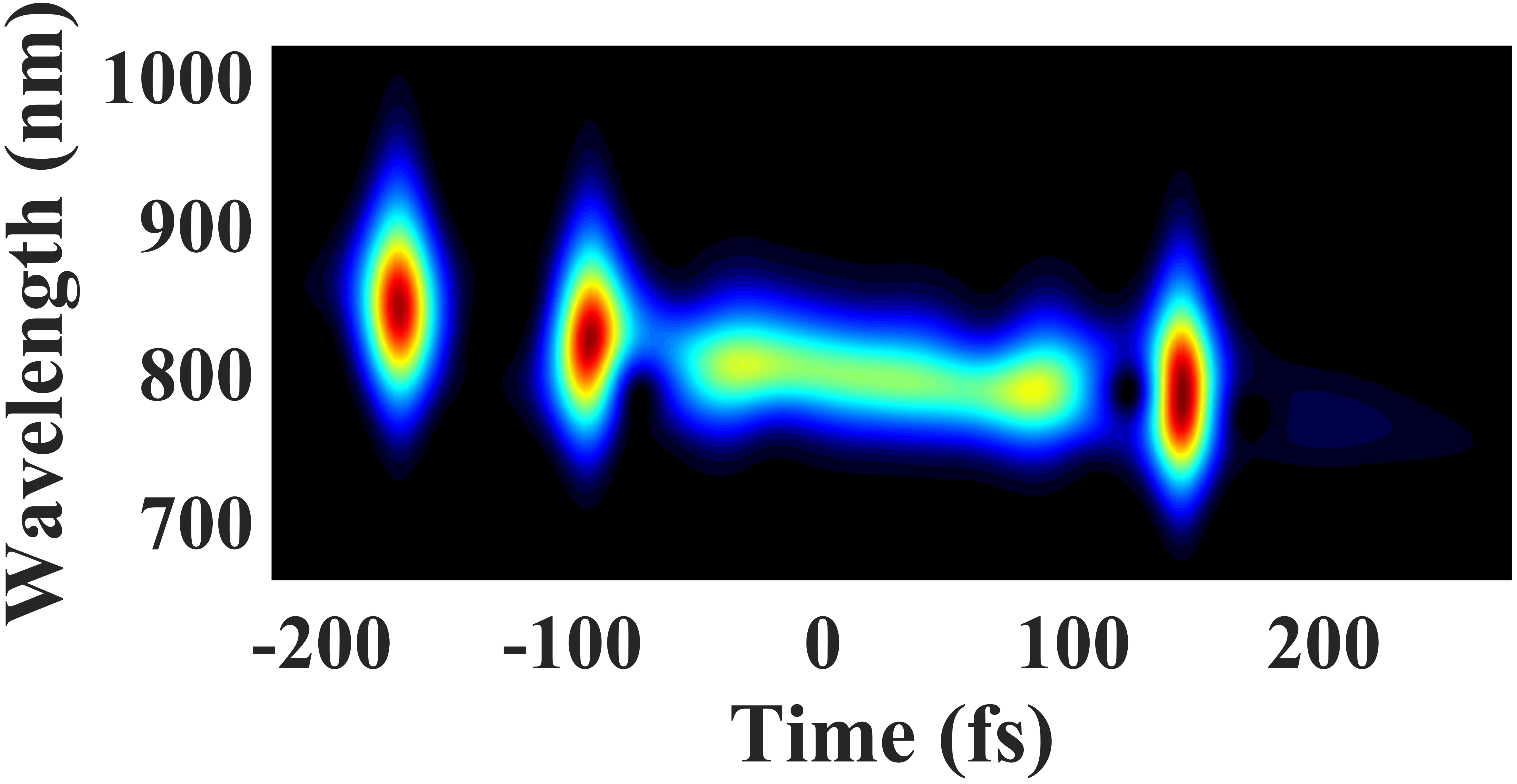}
         \caption{}
        \label{fig:6c}
    \end{subfigure}
    \begin{subfigure}[b]{0.315\textwidth}
        \includegraphics[width=\textwidth]{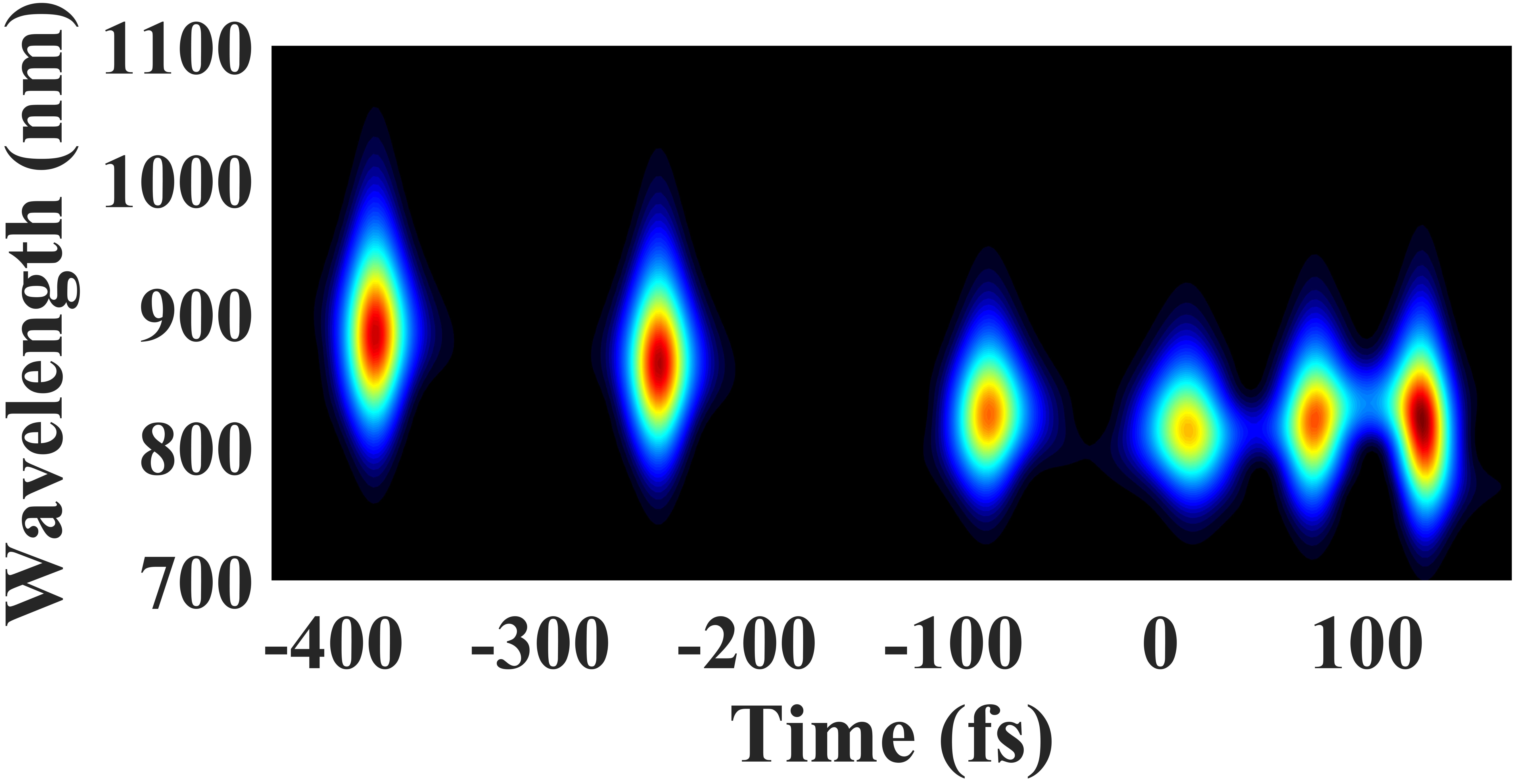}
         \caption{}
        \label{fig:6d}
    \end{subfigure}
  \begin{subfigure}[b]{0.315\textwidth}
        \includegraphics[width=\textwidth]{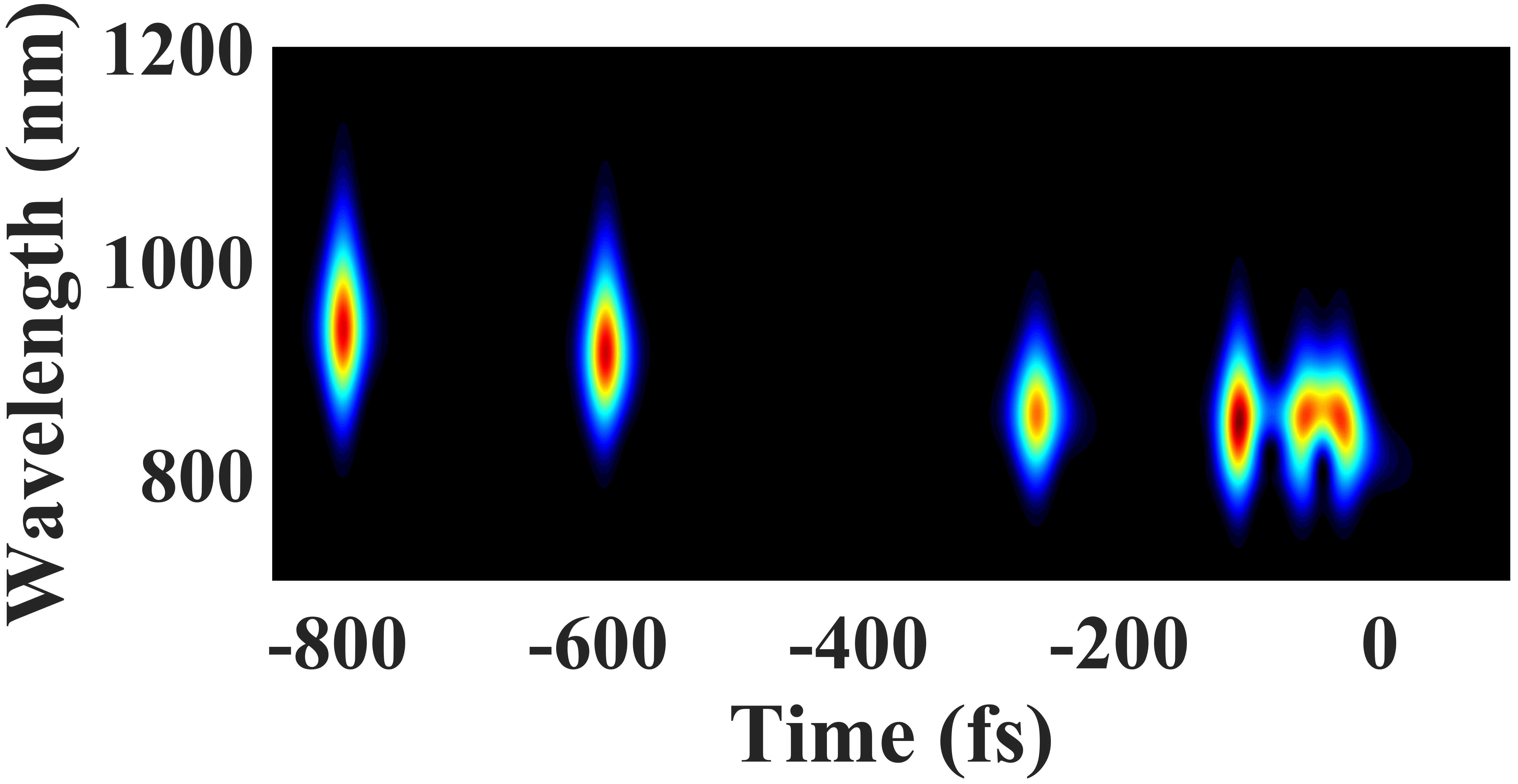}
         \caption{}
        \label{fig:6e}
    \end{subfigure}
\begin{subfigure}[b]{0.315\textwidth}
        \includegraphics[width=\textwidth]{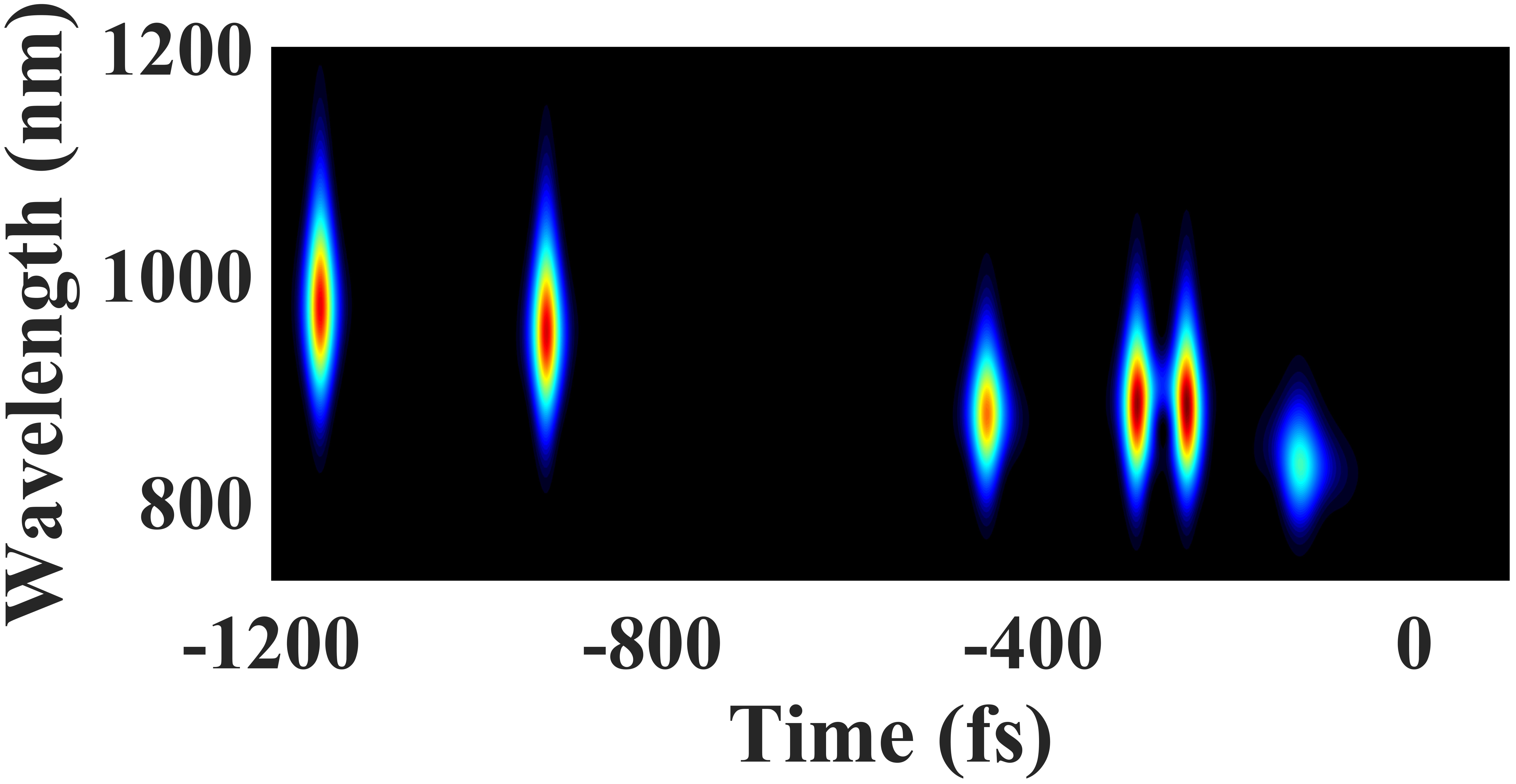}
         \caption{}
        \label{fig:6f}
    \end{subfigure}
   
    \caption{Output spectrograms of the system with $L_{1} = 15 mm$ initiated by the input pulse with peak power 0.25 MW. (a) At $L = 15 mm$ (the output of the SF segment); (b) at $L = 30 mm$; (c) at $L = 40 mm$; (d) at $L = 75 mm$; (e) at $L = 120 mm$; (f) at $L = 150 mm$.}
    
    \label{fig:frog}
\end{figure}

\section{Conclusions} \label{sect:Canclu}

In this work, we propose a method to generate multiple ultrashort temporal solitons by a pulse propagating in a composite waveguide consisting of two segments with opposite signs of cubic refractive nonlinearity, while the dispersion is normal in both segments. Systematic simulations of the corresponding generalized NL Schr\"{o}dinger equation demonstrate that pairs of temporal solitons are generated symmetrically with respect to the central region of the pulse, and as they propagate more solitons are generated from the leading and trailing edges until the central region of the pulse creating optical scenarios able to observe soliton collision. The physical process that enables the soliton generation is MTC (Multiple Temporal Compression), which may be obtained in systems composed of successive pairs of SF (self-focusing) and SDF (self-defocusing) materials. This may be considered as a scheme of \emph{nonlinearity management}, i.e., a chain of alternating SF and SDF segments with a common GVD value \cite{towers2002stable, jisha2019generation}.

The multi-soliton dynamics considered in this work can be extended by producing \emph{Newton's cradle} already investigated in \cite{driben2013newton}, i.e., propagation of collision waves in a multi-soliton chain.

Also, we demonstrated, for the first time, that the MTC process provides a method to generate ultrashort temporal solitons from a single input pulse, where the number of generated solitons can be controlled by the input peak power. This phenomenon is hard to be observed in a single waveguide under anomalous dispersion regime due to the energy limitation once a high energy concentration is localized in the first generated soliton around the central peak of the pulse.  

Exploitation of the MTC process is a powerful method for controllable generation of multiple solitons due to the energy redistribution among them. This approach was already exploited for generation of spatial solitons \cite{jisha2019generation} but investigation of a composite sample like the one studied here was not considered before. The main advantage of the present approach is the possibility to control the number of generated solitons, which is manageable due to the energy redistribution. This point will be further addressed in a separate work, where a detailed comparison between the MTC process and high-order soliton fission in a PCF fiber will be reported. Another relevant direction for further studies is implementation of MTC for generation of multiple dark solitons and subsequently addressing collisions between them.

\section{Acknowledgments} \label{sect:Acknowl}

This work was supported by the Brazilian agencies Conselho Nacional de Desenvolvimento Científico e Tecnol\'ogico - CNPq (Grant: 431162/2018-2 and the National Institute of Photonics (INCT) program - Grant: 465.763/2014), Funda\c{c}\~ao de Amparo \`a Ci\^encia e Tecnologia do Estado de Pernambuco (FACEPE), and the doctoral scholarship of A.C.A. Siqueira was provided by the CNPq. The work of B.A.M. was supported, in part, by the Israel Science Foundation through grant No. 1695/22.

\end{document}